\documentclass[a4paper,11pt]{article}
\usepackage{amsmath,mathtools,bm,amssymb,color}

\usepackage{graphicx}
\usepackage{subcaption}
\usepackage{jheppub} 
\usepackage{xspace}
\usepackage{booktabs}
\usepackage{array}
\usepackage{multirow}
\usepackage{adjustbox}
\usepackage{xcolor}
\usepackage[utf8]{inputenc}

\allowdisplaybreaks

\DeclareRobustCommand{\NNLOJET}{NNLO\scalebox{.8}{JET}\xspace}

\def\be{\begin{equation}}
\def\ee{\end{equation}}
\def\ba{\begin{eqnarray}}
\def\ea{\end{eqnarray}}

\title{NNLO QCD corrections in full colour for jet production observables at the LHC}

\author{X.\,Chen$^{a,b}$, T.\,Gehrmann$^{c}$, E.W.N.\,Glover$^d$, A.\,Huss$^e$, J.\,Mo$^c$}

\affiliation{
$^a$Institute for Theoretical Physics, Karlsruhe Institute of Technology, 76131 Karlsruhe, Germany\\
$^b$Institute for Astroparticle Physics, Karlsruhe Institute of Technology,\\ 76344 Eggenstein-Leopoldshafen, Germany\\
$^c$Physik-Institut, Universit\"at Z\"urich, Winterthurerstrasse 190, 8057 Z\"urich, Switzerland\\
$^d$Institute for Particle Physics Phenomenology, Department of Physics, University of Durham, Durham, DH1 3LE, UK\\
$^e$Theoretical Physics Department, CERN, 1211 Geneva 23, Switzerland}

\emailAdd{xuan.chen@kit.edu}
\emailAdd{thomas.gehrmann@uzh.ch}
\emailAdd{e.w.n.glover@durham.ac.uk}
\emailAdd{alexander.huss@cern.ch}
\emailAdd{jmo@physik.uzh.ch}

\abstract{
Calculations for processes involving a high multiplicity of coloured particles often employ a leading colour approximation, where only the leading terms in the expansion of the number of colours $N_c$ and the number of flavours $n_f$ are retained. This approximation of the full colour result is motivated by the $1/N_c^2$ suppression of the first 
subleading terms and by the increasing complexity of including subleading colour contributions to the calculation. In this work, we present the calculations using the antenna subtraction method in the \NNLOJET framework for the NNLO QCD corrections at full colour for several jet observables at the LHC. The single jet inclusive cross section is calculated doubly differential in transverse momentum and absolute rapidity and compared with the CMS measurement at 13\,TeV. A calculation for dijet production doubly differential in dijet mass and rapidity difference is also performed and compared with the ATLAS 7\,TeV data. Lastly, a triply differential dijet cross section in average transverse momentum, rapidity separation and dijet system boost is calculated and compared with the CMS 8\,TeV data. The impact of the subleading colour contributions to the leading colour approximation is assessed in detail for all three types of observables and as a function of the jet cone size. The subleading colour contributions
play a potentially sizable role in the description of the triply differential distributions, which probe kinematical configurations that are not easily accessed by any of the other observables.}

\keywords{Hadron Colliders, QCD Phenomenology, Jets, NNLO Corrections}
\preprint{{\raggedleft%
			ZU-TH 11/22\\
			KA-TP-07-2022\\
			IPPP/22/20\\
			P3H-22-037\\
			CERN-TH-2022-067\\
			}}

\begin{document}
\maketitle
\flushbottom

\section{Introduction}
The calculation of higher-order perturbative QCD corrections to collider observables proceeds through the evaluation of all real and virtual 
subprocess contributions at the desired order. Infrared divergences are typically present in each subprocess and only appropriate sums
of real and virtual contributions yield infrared-finite results, thereby requiring an infrared subtraction method to extract and recombine singular terms among different subprocesses. Substantial progress on the automation of one-loop virtual
corrections~\cite{Ellis:2011cr,Buccioni:2019sur} and their integration into multi-purpose event generator
programs~\cite{Alioli:2010xd,Alwall:2014hca,Kallweit:2015dum,Bellm:2015jjp}
now enables next-to-leading order (NLO) QCD calculations for processes of arbitrary multiplicity and associated infrared-safe observables.

At next-to-next-to-leading order (NNLO) and beyond, QCD calculations are performed on a process-by-process basis.
Several subtraction methods have been developed for NNLO 
calculations~\cite{Catani:2007vq,Gehrmann-DeRidder:2005btv,Currie:2013vh,Czakon:2010td,Boughezal:2011jf,Gaunt:2015pea,DelDuca:2016ily,Caola:2017dug} 
and are being applied to 
a range of hadron collider processes, reviewed in detail in~\cite{Heinrich:2020ybq}. 
Up to now, these calculations have been limited to low final state multiplicities 
(mostly $2\to 2$ processes) by
two major constraints: the availability of two-loop virtual corrections to the relevant scattering amplitudes, and the sheer computational complexity 
of the real radiation corrections and the associated infrared subtraction. Despite substantial progress being made on both fronts, NNLO calculations for processes involving three or more
particles in the final state remains very demanding in terms of computation time and complexity. 

By structuring the corrections at a given perturbative order into numerically dominant and subdominant pieces, a substantial gain in terms of 
computational efficiency 
can be achieved. A natural approach to identify the dominant QCD contributions is through an expansion in the number of colours $N_c$ that further facilitates a decomposition into separately gauge-invariant contributions.
The leading-colour contributions to a given process take a particularly simple form at each order: virtual loop corrections 
contain only planar diagrams~\cite{tHooft:1973alw}, and real radiation corrections are obtained by summing squares of colour-ordered 
Feynman amplitudes~\cite{Mangano:1990by}, thereby discarding all interference contributions, rendering their evaluation considerably more
efficient and simplifying their 
infrared subtraction in certain schemes~\cite{Gehrmann-DeRidder:2005btv,Currie:2013vh}. 

Most recently, first results for NNLO corrections to $2\to 3$
hadron collider processes were obtained for three-photon production~\cite{Chawdhry:2019bji,Kallweit:2020gcp}, 
diphoton-plus-jet production~\cite{Chawdhry:2021hkp} 
and three-jet production~\cite{Czakon:2021mjy}. All these works include subleading colour terms only partially: the two-loop 
virtual QCD corrections to the scattering amplitudes are evaluated in the leading colour
approximation~\cite{Abreu:2020cwb,Chawdhry:2020for,Agarwal:2021grm,Chawdhry:2021mkw,Abreu:2021oya} while all subleading colour terms are included for the real radiation subprocesses. With all massless two-loop five-point integrals being known~\cite{Chicherin:2018old} and available in a format suitable for numerical 
evaluation~\cite{Chicherin:2020oor}, the previously missing full-colour results for virtual two loop scattering amplitudes no longer pose a fundamental 
restriction; they are now starting to become 
available~\cite{Agarwal:2021vdh,Badger:2021imn} and can be expected to be included in cross section evaluations in due course.  
A leading colour approximation has also been used previously for NLO corrections to high multiplicity processes, for example for 
$W+4$-jet and $W+5$-jet final states~\cite{Berger:2010zx,Bern:2013gka}, to enhance the computational efficiency 
and convergence of the calculations. A weighted sampling over the different colour levels is done routinely in many multi-purpose event 
generator programs. 

It is the purpose of the current paper to quantify the quality of the leading-colour approximation at NNLO for 
hadron collider dijet production and related observables, such as the single jet inclusive cross section. These processes were computed 
previously 
to leading colour \cite{Currie:2016bfm,Currie:2017eqf,Gehrmann-DeRidder:2019ibf} using the antenna subtraction 
method~\cite{Gehrmann-DeRidder:2005btv,Currie:2013vh,Daleo:2006xa} 
in the \NNLOJET framework. These leading-colour results are being used extensively
for precision phenomenology, in particular in view of the determination of parton distributions at NNLO~\cite{AbdulKhalek:2020jut}.
 A subsequent calculation, based on a residue subtraction 
technique~\cite{Czakon:2010td}, including all colour levels for the single jet inclusive cross section~\cite{Czakon:2019tmo} 
at one specific jet resolution parameter reported little to no numerical evidence 
for the subleading colour contributions. In the current work, we extend the \NNLOJET calculation of hadron collider jet production  
to all colour levels and quantify the impact of the subleading colour contributions at NNLO for a variety of observables. 
While the details of the implementation are documented in a companion paper~\cite{Chen:2022clm}, the current work focuses on the numerical 
results and their phenomenological implications. 

The paper is structured as follows. In Section~\ref{sec:theory}, we briefly describe the calculation of the NNLO QCD corrections 
to jet production processes and define the accounting of leading colour and subleading colour contributions. Section~\ref{sec:single}
quantifies the relative impact of leading and subleading colour contributions to the single jet inclusive cross section at NNLO, 
investigating different jet resolutions and examining the contributions from different partonic initial states. Similar considerations are 
repeated for dijet observables in Sections~\ref{sec:dijet} and~\ref{sec:3d_2jet}. The numerical impact of the subleading colour contributions 
varies strongly between the different observables, being most pronounced in the triple differential dijet distributions described in Section~\ref{sec:3d_2jet}. 
In our conclusions in Section~\ref{sec:conc}, we discuss the pattern of subleading colour effects that we observe in the different jet cross sections and elaborate on 
its possible relation to the kinematical definitions of the observables.

\section{Jet production cross sections at NNLO QCD}
\label{sec:theory}

The basic Born-level jet production process at hadron colliders is $2\to 2$ parton scattering, which yields a final state containing two jets that are 
back-to-back in transverse momentum. Jets are reconstructed using a jet algorithm (at LHC mainly with the anti-$k_T$ algorithm~\cite{Cacciari:2008gp}), 
and characterised by their transverse momentum $p_T$, rapidity $y$ and azimuthal angle $\phi$. Jets in an event are ordered in 
decreasing transverse momentum and kinematical acceptance cuts are applied in $y$ and $p_T$. 
Several types of observables can be derived from this Born-level process, broadly distinguished into two classes. 
Single-jet inclusive cross sections receive contributions from each jet in an event that is within the kinematical acceptance, such that each event 
can contribute multiple times. Inclusive dijet cross sections receive contributions from all events containing at least two jets within the kinematical 
acceptance, and their kinematics is reconstructed based on the two leading jets only, such that each event contributes exactly once to the cross section. 

The QCD description of single jet inclusive and of dijet cross sections is based on the same $2\to 2$ parton-level processes, corrected to the 
desired order in perturbation theory. The calculation of the QCD corrections is performed in the form of a parton-level event generator, 
which provides the full kinematical information for each event, allowing  single jet inclusive and dijet observables to be reconstructed. The 
higher order contributions amount to real and virtual corrections to the underlying Born-level subprocesses. Only the sum of real and 
virtual corrections is infrared finite, such that a subtraction method is required to isolate and recombine infrared singular terms 
from individual subprocess contributions, allowing them to be implemented into the parton-level event generator. 

Our calculation of NNLO QCD corrections to jet production processes at hadron colliders is based on the antenna subtraction 
method~\cite{Gehrmann-DeRidder:2005btv,Currie:2013vh,Daleo:2006xa} and implemented in the \NNLOJET parton-level event generator. 
At variance with all other processes in \NNLOJET, which are all computed in full colour to NNLO, hadron collider 
jet production was previously implemented in \NNLOJET 
retaining only the leading colour terms at NNLO~\cite{Currie:2016bfm,Currie:2017eqf,Gehrmann-DeRidder:2019ibf}. 
The NNLO QCD corrections receive contributions 
from tree-level $2\to 4$ parton processes~\cite{Mangano:1990by},  from one-loop $2\to 3$ parton 
processes~\cite{Bern:1993mq,Bern:1994fz,Kunszt:1994nq} and from two-loop $2\to 2$ parton 
processes~\cite{Glover:2001af,Anastasiou:2000kg,Anastasiou:2000ue,Bern:2002tk,Bern:2003ck,DeFreitas:2004kmi}. 
The antenna subtraction for these subprocesses can be constructed from the known set of antenna functions in 
final-final~\cite{Gehrmann-DeRidder:2005btv},
intial-final~\cite{Daleo:2009yj} and initial-initial kinematics~\cite{Gehrmann:2011wi,Gehrmann-DeRidder:2012too}. The subtraction terms 
at subleading colour levels must account for complicated interference structures related to the single and double soft singular behaviour of 
the real radiation contributions. 
For hadron collider jet production, these were treated previously in antenna subtraction only for purely gluonic subprocesses~\cite{Currie:2013dwa,Chen:2022ktf}.
 Their construction for all parton-level processes 
are described in detail in a companion paper~\cite{Chen:2022clm}.

\subsection{Leading and subleading colour contributions}
Already at leading order, all parton-level initial states (gluon-gluon, quark-gluon, quark-(anti-)quark in all flavour combinations) 
contribute to jet production cross sections. With an overall factor $(N_c^2-1)$ from all Born-level $2\to 2$ squared matrix elements, one finds only a single colour structure in the four-gluon and four-quark
 squared matrix elements, and a leading ($N_c^0$) and subleading ($1/N_c^2$) colour structure for the two-quark-two-gluon
 squared  matrix elements. 
 Depending on the kinematical crossing under consideration, the respective Born-level subprocess cross sections contain two 
 factors of $1/N_c$ or $1/(N_c^2-1)$ 
 to account for initial-state quark or gluon colour averages and can contain a factor $n_f$ counting the number of flavours in the final state. 
 
 The notion of leading colour (LC) or subleading colour  (SLC) corrections (with their sum being the full colour, FC, contribution) at NLO and NNLO 
 is defined for each parton-level subprocess relative to the respective Born level, keeping the exact Born-level normalization factors (discarding the
 subleading colour contributions to  two-quark-two-gluon
 squared  matrix elements). 
 
 The NLO corrections to a given Born-level process amount to relative factors $N_c$, $n_f$ and $1/N_c$, where the former two constitute the LC correction, and the third yields the SLC correction. At NNLO, LC consists of $N_c^2$, $n_f N_c$ and $n_f^2$, and SLC amounts to $N_c^0$, $n_f/N_c$, $1/N_c^2$. 
For all processes that originate from Born-level two-quark-two-gluon crossings, the NLO corrections will also contain SLC 
contributions (from corrections to the SLC Born matrix elements) 
proportional to  $n_f/N_c^2$ and $1/N_c^3$ relative to the leading colour Born process, and NNLO contains further 
SLC contributions at $n_f^2/N_c^2$, $n_f/N_c^3$ and $1/N_c^4$. 

Throughout this paper, we will discuss the impact of leading and subleading colour contributions both on the cross section at a given perturbative order and 
on the perturbative contribution at this order. To distinguish these, we use the following notation:
\begin{eqnarray}
\mathrm{d}\sigma_{\mathrm{NLO}} &=& \mathrm{d}\sigma_{\mathrm{LO}} +  \mathrm{d}\delta\sigma_{\mathrm{NLO}} \,,\nonumber\\
 \mathrm{d}\sigma_{\mathrm{NNLO}} &=& \mathrm{d}\sigma_{\mathrm{LO}} +  \mathrm{d}\delta\sigma_{\mathrm{NLO}}  +  \mathrm{d}\delta\sigma_{\mathrm{NNLO}}\,.
 \end{eqnarray}
All terms in the above equations are to be understood at full colour, and the perturbative fixed-order contributions are decomposed into LC and SLC 
pieces:
 \begin{eqnarray}
     \mathrm{d}\delta\sigma_{\mathrm{(N)NLO}} & = &  \mathrm{d}\delta\sigma_{\mathrm{(N)NLO}}^{\mathrm{LC}} + \mathrm{d}\delta\sigma_{\mathrm{(N)NLO}}^{\mathrm{SLC}}\,.
  \end{eqnarray}
The LC approximation on the cross section is then applied only to the coefficient from the highest perturbative order (and not to all coefficients), 
to quantify the effect of the previously 
uncalculated terms:
\begin{eqnarray}
\mathrm{d}\sigma_{\mathrm{NLO}}^{\mathrm{LC}}  &=& \mathrm{d}\sigma_{\mathrm{LO}} +  \mathrm{d}\delta\sigma_{\mathrm{NLO}}^{\mathrm{LC}}  \,,\nonumber\\
 \mathrm{d}\sigma_{\mathrm{NNLO}}^{\mathrm{LC}}  &=& \mathrm{d}\sigma_{\mathrm{LO}} +  \mathrm{d}\delta\sigma_{\mathrm{NLO}}  +  \mathrm{d}\delta\sigma_{\mathrm {NNLO}}^{\mathrm{LC}} \,.
 \end{eqnarray}

\section{Single jet inclusive cross section}
\label{sec:single}
In this section we study the most basic jet production observable from hadron-hadron colliders, the single jet inclusive cross section
\begin{align*}
\mathrm{d}\sigma(p+p \to j+X) \,,
\end{align*}
which is obtained by considering all jets in any given event with at least one jet.
It is inclusive over additional radiation, such that any additional jet from multi-jet events within the acceptance is taken into account.
Measurements of the single jet inclusive cross section have been performed already at early hadron collider 
experiments~\cite{UA1:1986okk,UA2:1982nvv}, followed by precision measurements at the Fermilab Tevatron~\cite{CDF:2008hmn,D0:2011jpq}.
The LHC experiments have studied single jet inclusive production over a wide kinematical range at various LHC collider 
energies~\cite{ALICE:2013yva,ATLAS:2010dmq,ATLAS:2013pbc,ATLAS:2017kux,ATLAS:2017ble,CMS:2012ftr,CMS:2015jdl,CMS:2016lna,CMS:2016jip,CMS:2021yzl}.

QCD corrections  to single jet inclusive production are known at NLO already for a long time~\cite{Ellis:1992en,Giele:1994gf,Giele:1994xd}. 
They are part of standard simulation codes~\cite{Nagy:2001fj,Gao:2012he}, 
have been matched to parton shower simulations~\cite{Alioli:2010xa,Hoeche:2012fm}
and are also complemented by NLO electroweak corrections~\cite{Dittmaier:2012kx,Campbell:2016dks,Frederix:2016ost}. NNLO QCD corrections 
were first computed by retaining only the leading-colour contributions~\cite{Currie:2016bfm,Currie:2017eqf,Gehrmann-DeRidder:2019ibf}, 
and subsequently to all colour-levels~\cite{Czakon:2019tmo}.

Owing to its definition through a sum over all jets in an event, the higher order corrections to the single jet inclusive cross section display certain 
pathological features~\cite{Dasgupta:2016bnd,Cacciari:2019qjx} in terms of stability and slow perturbative convergence. 
An extensive theoretical study of this observable at NNLO was performed in \cite{Currie:2018xkj}. In particular, the single jet inclusive cross section was calculated doubly differential in the transverse jet momentum $p_T$ and absolute rapidity $|y|$, and compared with the CMS measurements at $\sqrt{s}=13\,\mathrm{TeV}$  \cite{CMS:2016jip}. These predictions included NNLO QCD corrections using the leading-colour approximation, defined as the leading $N_c$ and $n_f$ terms for all partonic channels. In this section we discuss the computation of these doubly differential single jet inclusive cross sections with \NNLOJET, where the predictions now include the NNLO QCD corrections at full colour.

\subsection{Calculational setup}
Improving upon the earlier leading-colour NNLO results~\cite{Currie:2018xkj}, the double differential 
single jet inclusive cross section for the CMS measurement at $\sqrt{s} = 13\,\mathrm{TeV}$ is now calculated at NNLO full colour. The calculational setup is largely the same as for the
LC-only calculation with only a few differences and is briefly summarized here. Jets are identified with the anti-$k_T$ jet algorithm~\cite{Cacciari:2008gp} using cone sizes of $R=0.4$ or $R=0.7$. 
The following fiducial cuts on the jet transverse momentum and absolute rapidity are applied:
\be
114 \, \mathrm{GeV} < p_T < 2000 \, \mathrm{GeV}, \quad\quad |y| \, < 4.7
\ee
Instead of the PDF set PDF4LHC15$\_$nnlo$\_$100~\cite{Butterworth:2015oua} which was used for the LC-only calculation, we now use 
NNPDF40$\_$nnlo$\_$as$\_$01180~\cite{Ball:2021leu} as the default PDF set for all LO, NLO and NNLO coefficients. LHAPDF~\cite{Buckley:2014ana} is used to evaluate the PDF sets and the strong couping constant $\alpha_s$. The single jet inclusive cross section is calculated as a function of $p_T$ for seven slices in absolute rapidity:
\be
|y|: [0, 0.5];\, [0.5, 1];\, [1, 1.5];\, [1.5, 2];\, [2, 2.5];\, [2.5, 3];\, [3.2,4.7],
\ee 
and the binning in $p_T$ corresponding to the CMS 13\,TeV jet measurement with 35 bins in the range
$114 \, \mathrm{GeV} < p_T < 2000 \, \mathrm{GeV}$. 

The theoretical uncertainties are obtained by using a seven-point scale variation of the renormalization scale $\mu_R$ and factorization scale $\mu_F$ by factors of 1/2 and 2 around their central value. The envelope of all the scale variations, with the additional constraint that $1/2 \le \mu_R / \mu_F \le 2$, then forms the theoretical uncertainty band. For the central value of $\mu_F$ and $\mu_R$ we use the scalar sum of the transverse momenta of all partons: $\hat{H}_T = \sum_{i \in \mathrm{partons}} p_{T,i}$, as this scale choice was found to be the most appropriate for this process in terms of convergence and stability of the perturbative predictions~\cite{Currie:2018xkj}.
\begin{figure}[t]
	\centering
	\includegraphics[width=0.49\textwidth]{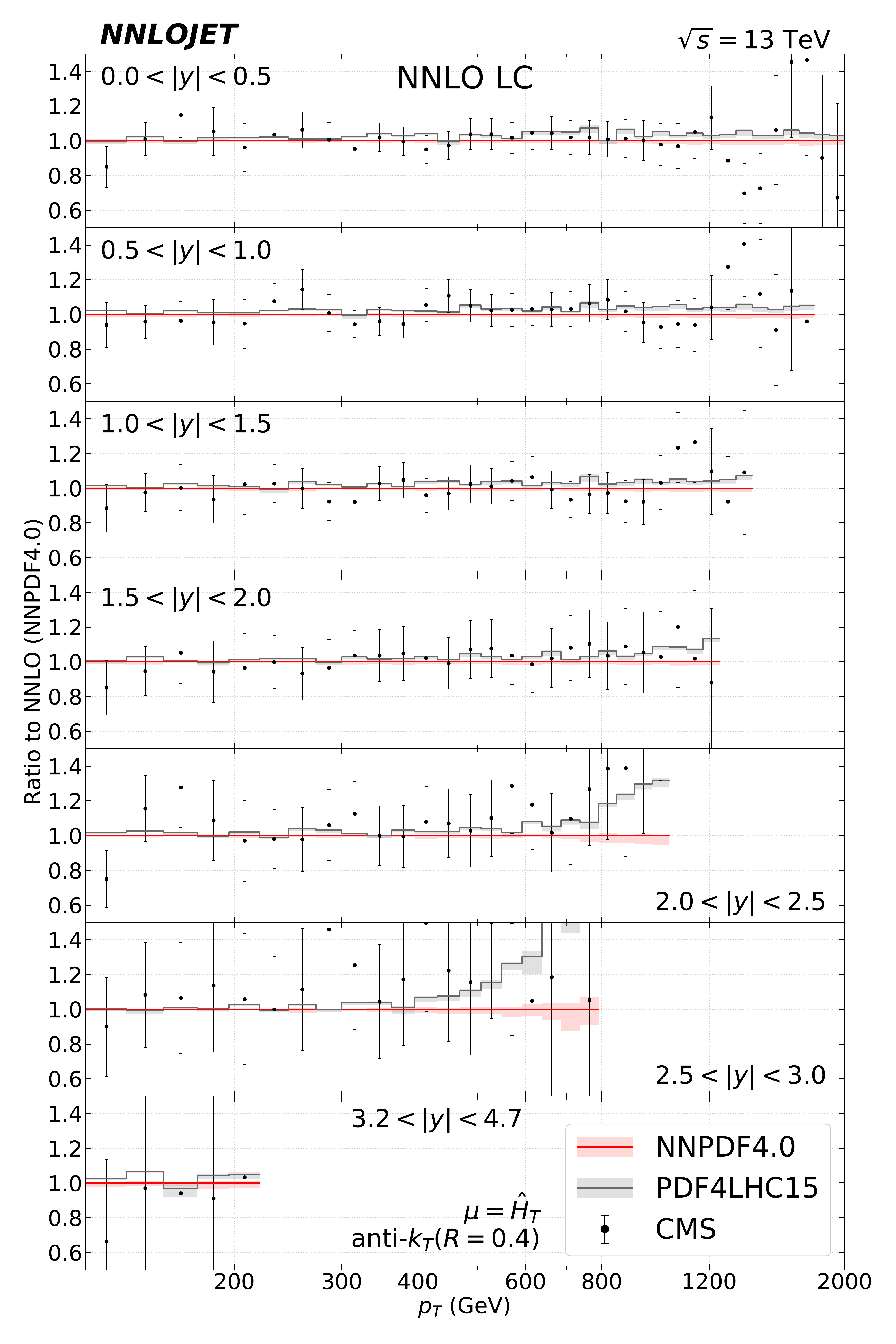}
         \includegraphics[width=0.49\textwidth]{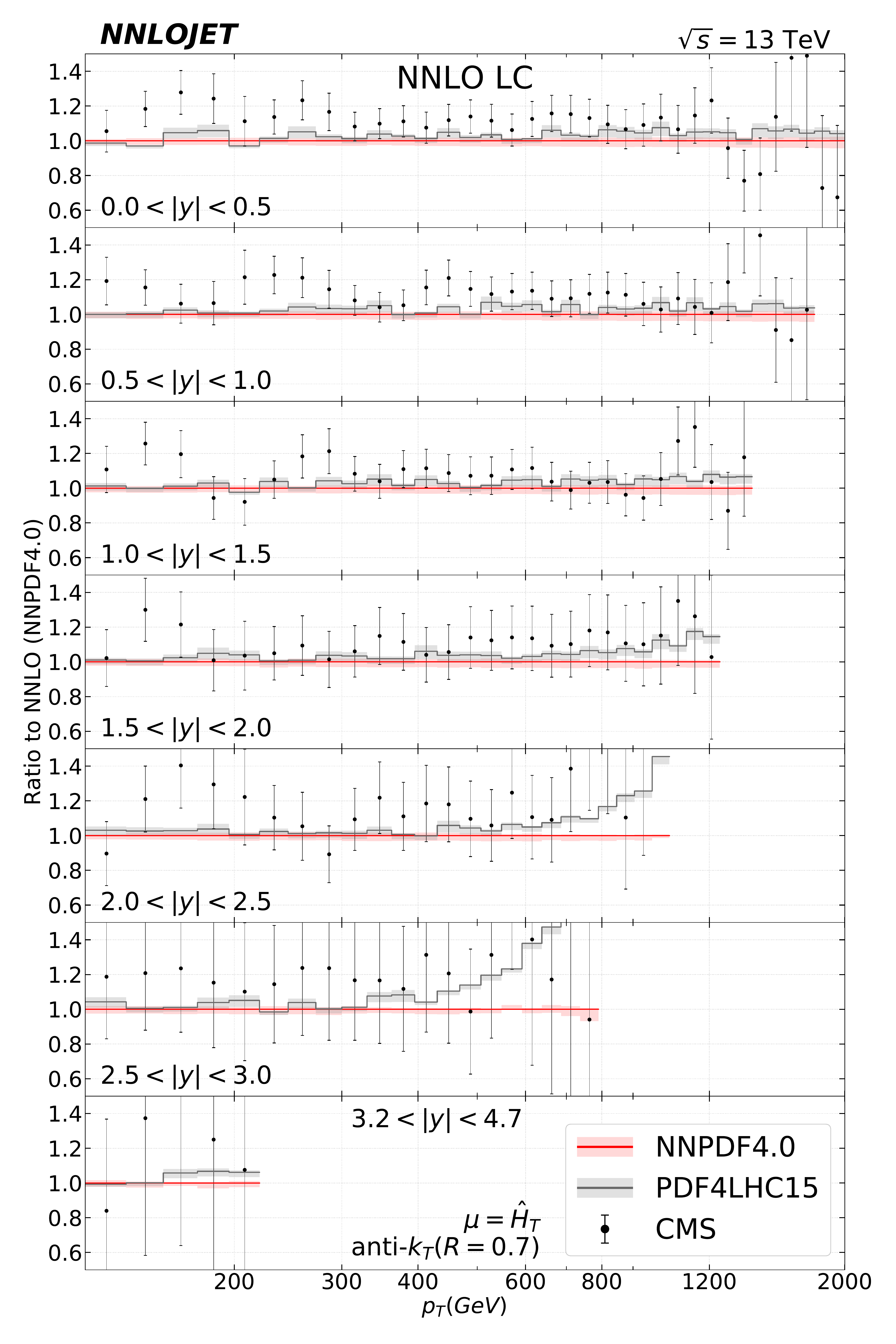}
	\caption{Comparison of LC-only NNLO  single jet inclusive distributions using PDF4LHC15$\_$nnlo$\_$100 (grey) 
	and NNPDF40$\_$nnlo$\_$as$\_$01180 (red) as the PDF for jet cone size $R=0.4$ (left) and $R=0.7$ (right).  } 
	\label{fig:PDFcomp}
\end{figure}

\subsection{PDF differences}  
Before examining the results of the full colour calculation, we first begin with an assessment of the differences in the distributions due to the different PDF sets used.
This comparison is performed at NNLO in leading-colour, to quantify potential changes due to PDF effects with respect to the earlier results~\cite{Currie:2018xkj}.
Data on  single jet inclusive cross sections are ingredients to the determination of PDFs from a global fit. They constrain in particular the shape of the 
gluon distribution~\cite{AbdulKhalek:2020jut}. Jet data from  LHC single jet inclusive measurements at 7\,TeV and 8\,TeV were newly included in
NNPDF4.0~\cite{Ball:2021leu}
and the difference in predictions between PDF4LHC15 and NNPDF4.0 highlights their impact. 

The LC approximation is computed using NNPDF4.0. In Figure~\ref{fig:PDFcomp}  a comparison is made between the distributions at LC-only NNLO. For both cone sizes the predictions overlap within scale uncertainties in the low $p_T$ region (with NNPDF4.0 results
being smaller than PDF4LHC results by 1 to 3\%), while at higher $p_T$ the prediction using 
NNPDF4.0
starts to decrease more rapidly than the PDF4LHC15
prediction, most notable for the higher absolute rapidity slices, where the uncertainty bands stop overlapping in the tail of the distribution. With large experimental errors in this region, both predictions remain consistent with the 
CMS data. 
\begin{figure}[t]
	\centering
	\includegraphics[width=0.49\textwidth]{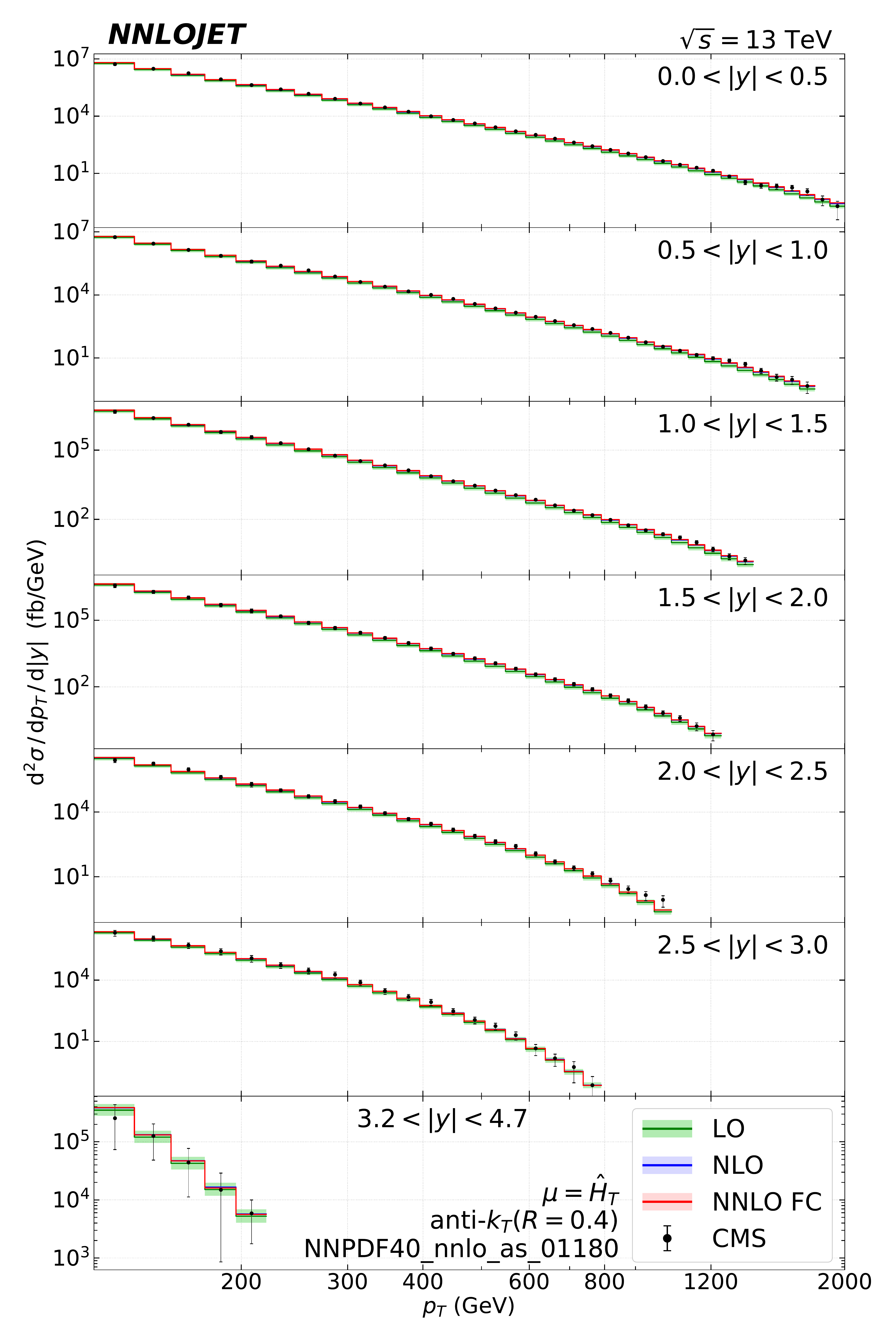} 
	\includegraphics[width=0.49\textwidth]{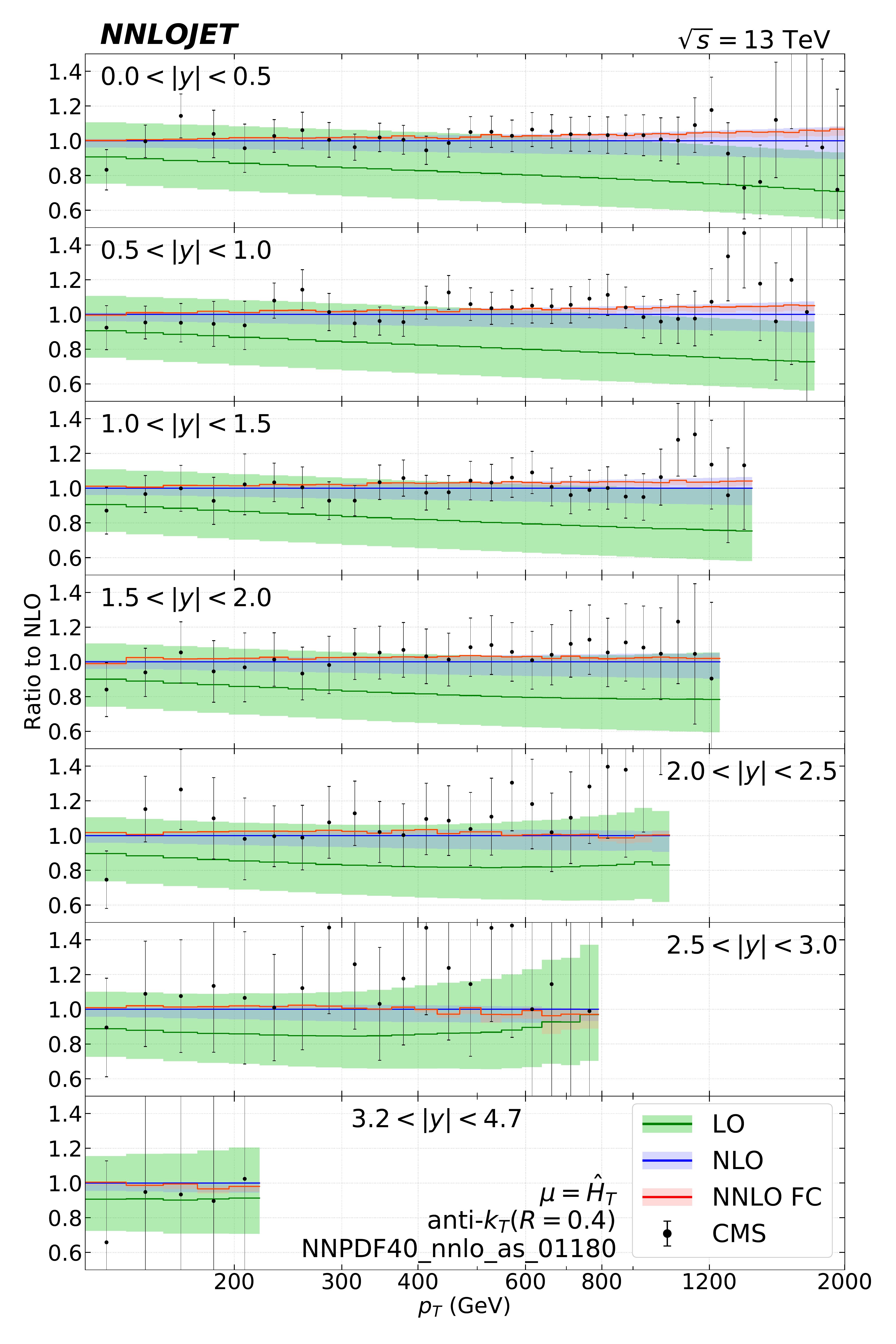}
	\caption{Double differential single jet inclusive cross sections as function of the jet $p_T$ for all rapidity slices with $R=0.4$ anti-$k_T$ jets (left), and normalized to the NLO result (right).  Predictions are compared to CMS data~\protect\cite{CMS:2016jip}.}
	\label{fig:CMS13_1j_R04}
\end{figure}

\subsection{Results}
\begin{figure}[t]
	\centering
	\includegraphics[width=0.49\textwidth]{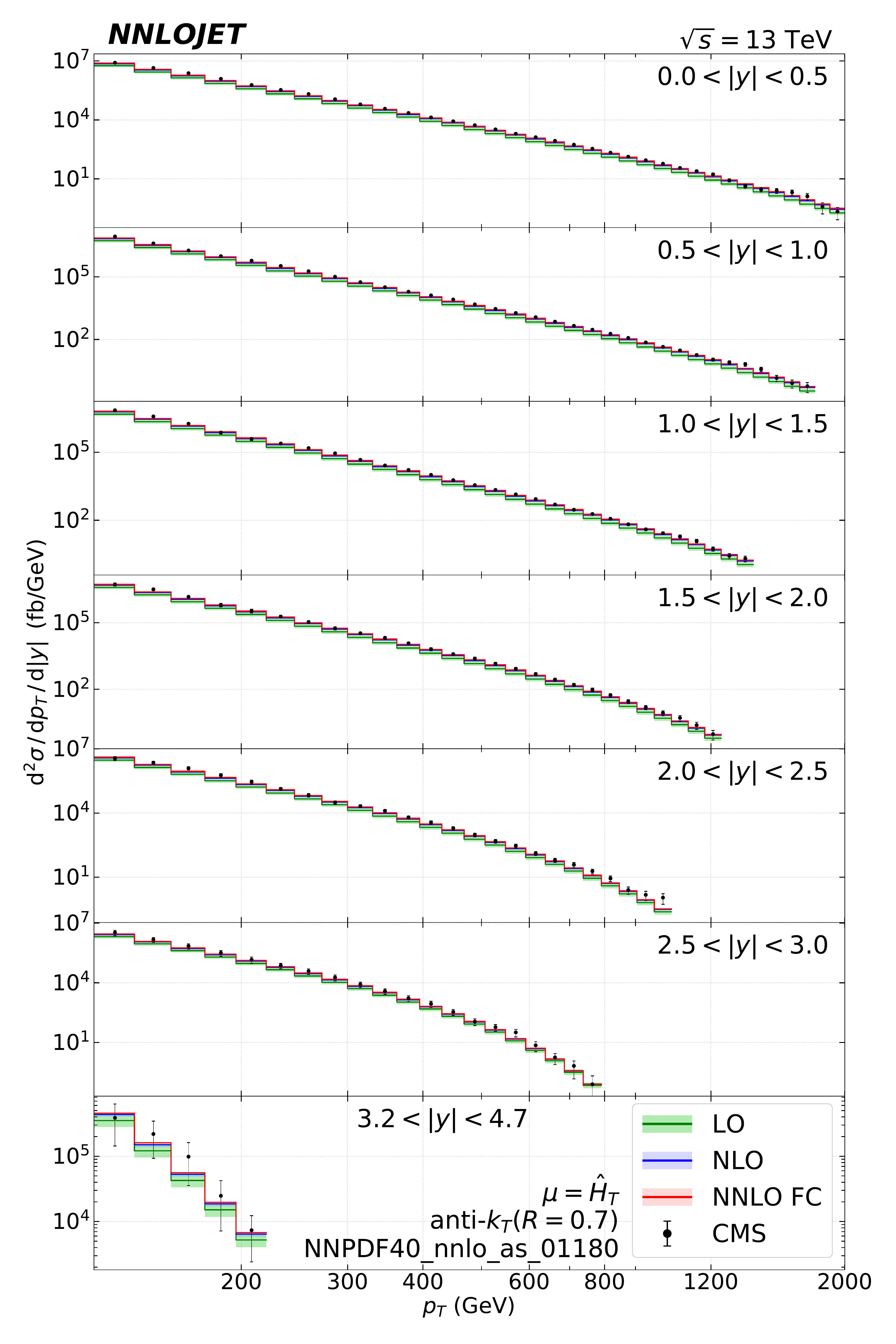} 
	\includegraphics[width=0.49\textwidth]{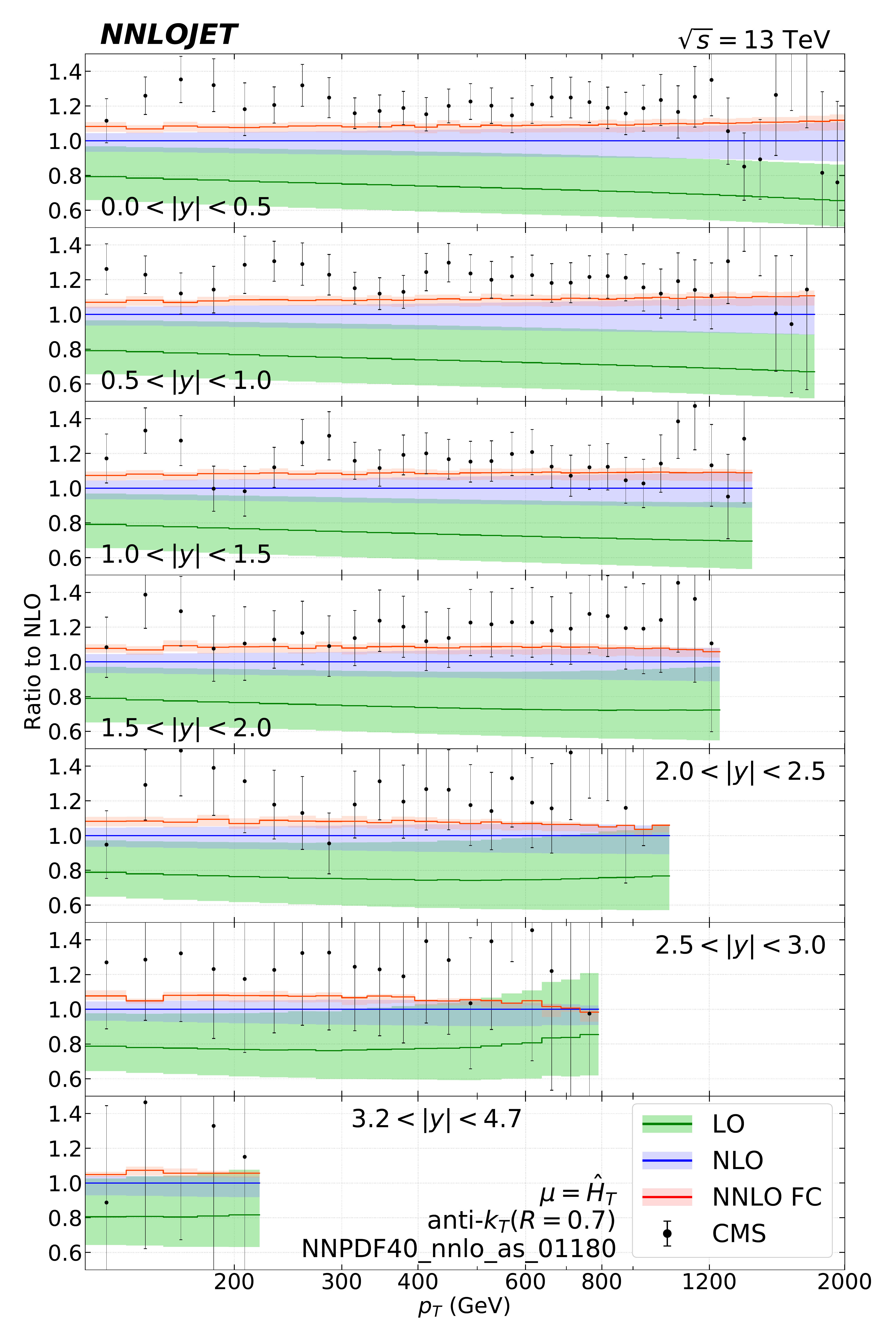}
	\caption{Double differential single jet inclusive cross sections as function of the jet $p_T$ for all rapidity slices with $R=0.7$ anti-$k_T$ jets (left), and normalized to the NLO result (right). Predictions are compared to CMS data~\protect\cite{CMS:2016jip}.}
	\label{fig:CMS13_1j_R07}
\end{figure}

In Figures~\ref{fig:CMS13_1j_R04} and \ref{fig:CMS13_1j_R07} the LO, NLO and NNLO full colour predictions for anti-$k_T$ jets with $R=0.4$ and $R=0.7$ respectively are compared with the CMS data. For $R=0.4$, we observe an excellent perturbative convergence, with modest NLO and very small NNLO corrections, each
prediction being fully contained within the scale uncertainty band of the preceding order. In the case of the larger cone size $R=0.7$, NLO and NNLO corrections are both larger, and the scale uncertainty bands of consecutive orders are merely overlapping with each other, with the central predictions typically outside the band from the preceding order.  For both cone sizes, already the NLO prediction leads to a significant reduction of the uncertainty band and better agreement with the data than for the LO prediction. The NNLO corrections lead to a further reduction of uncertainty in both cases,
and slightly improve the description of the experimental data. 

The NLO corrections are positive throughout the entire $p_T$ range for both values of the cone size. In contrast, the NNLO corrections for $R=0.4$ 
are slightly negative for small $p_T$, they change sign with increasing $p_T$ and and display a small but steady growth with $p_T$. They remain 
negative in the most forward rapidity bins. At $R=0.7$, the NNLO corrections are positive throughout, and nearly independent on $p_T$
in each given rapidity bin. The smallness of the full-colour NNLO corrections and their sign-change must be kept in mind in the following, when 
we compare leading-colour and full-colour NNLO coefficients for $R=0.4$. With positive NNLO corrections throughout at $R=0.7$, such comparisons are more meaningful in a quantitative manner.   
  \begin{figure}[t]
	\centering
	\includegraphics[width=0.49\textwidth]{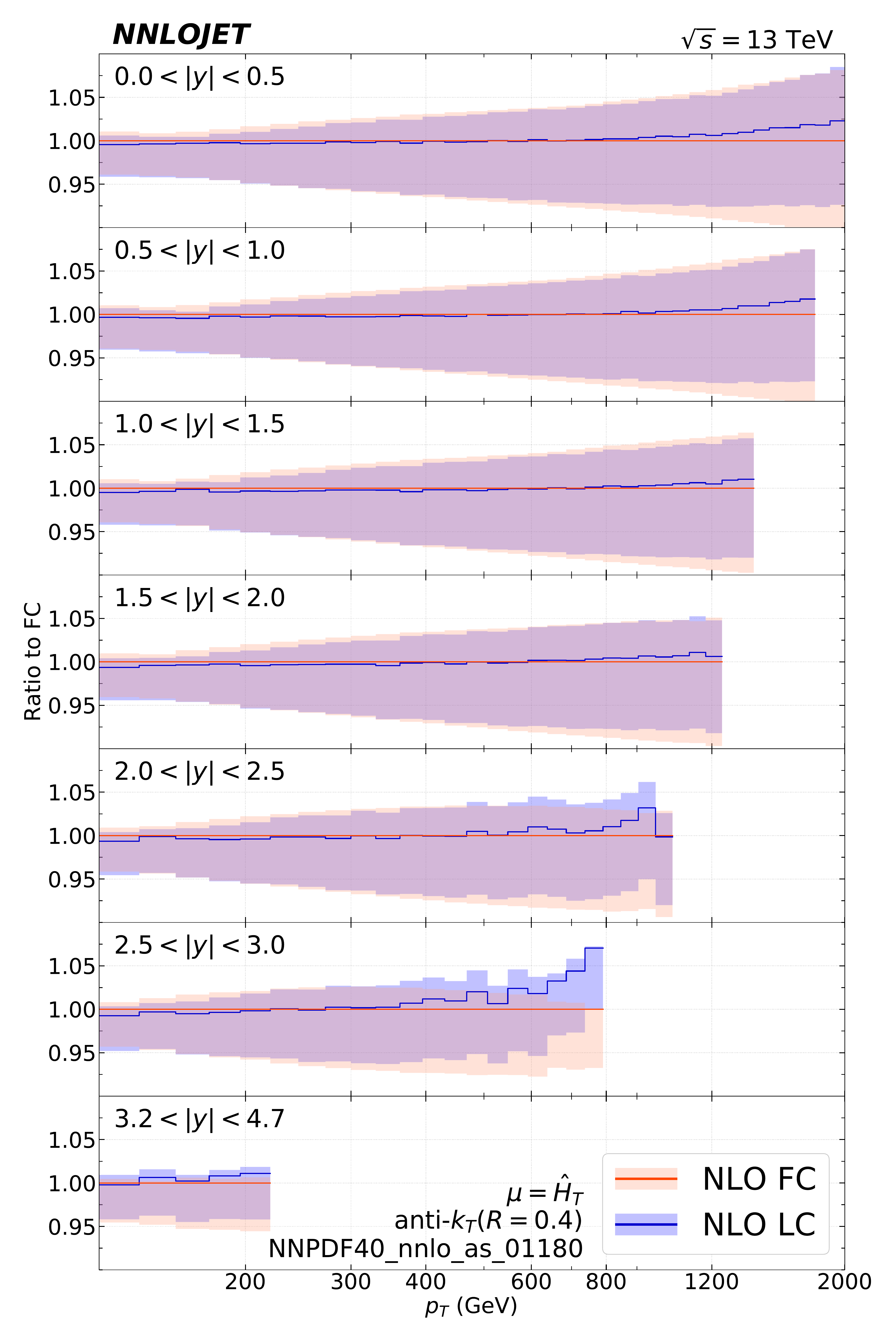}
	\includegraphics[width=0.49\textwidth]{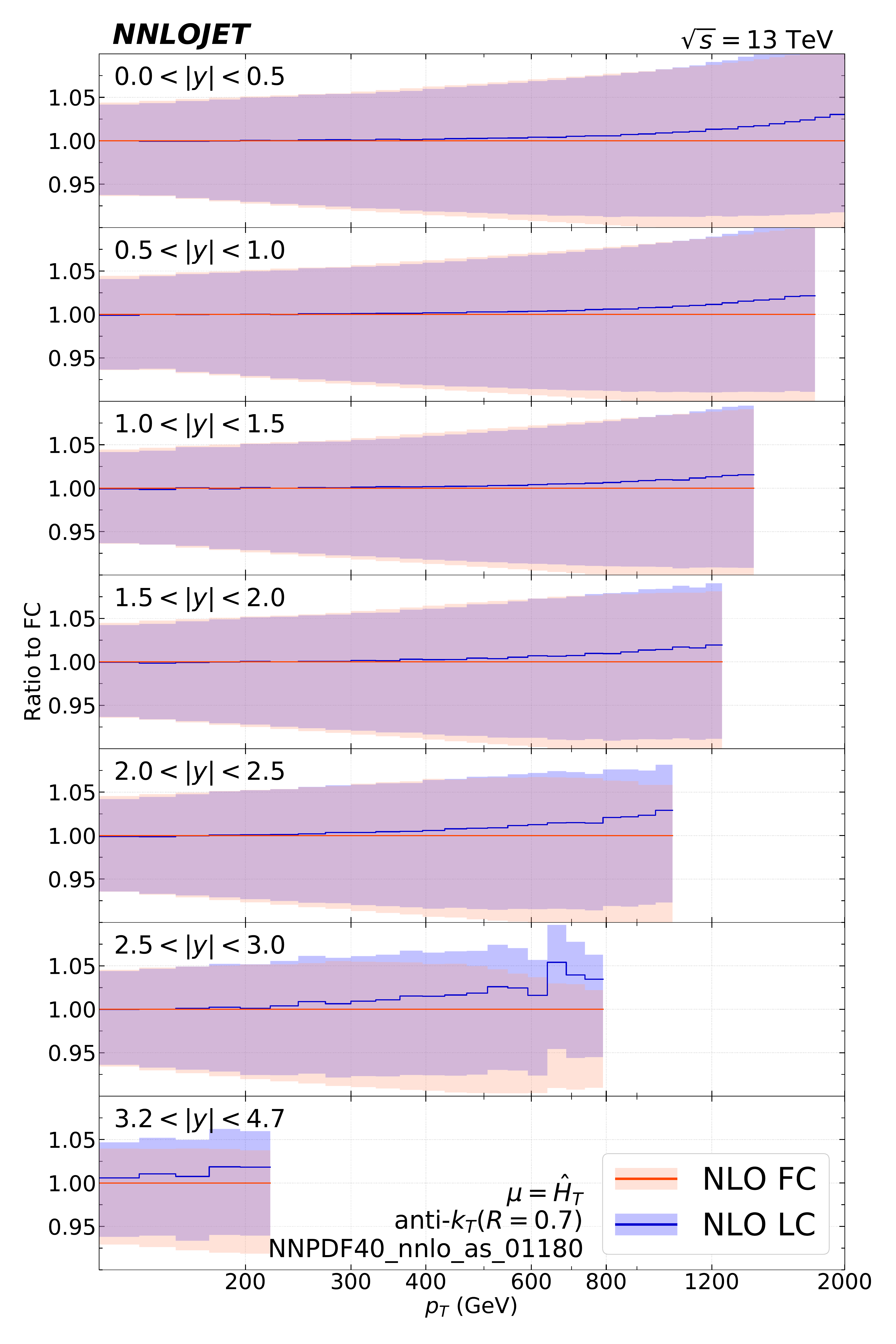}
	\caption{NLO predictions at LC (blue) and FC (red) predictions of the $R=0.4$ (left) and $R=0.7$ (right) single jet inclusive cross section as function of $p_T$ for all the considered rapidity slices, normalized to the FC prediction.}
	\label{fig:CMS13_1j_LCFC_NLO}
\end{figure}

\subsection{Subleading colour contributions}
Prior to investigating the numerical impact of the SLC contributions at NNLO, we first quantify their size at NLO, thereby
obtaining a first impression on their potential relevance. 
Figure~\ref{fig:CMS13_1j_LCFC_NLO} compares the NLO predictions for the single-jet inclusive cross sections at LC and FC, where the 
LC truncations is applied only to the NLO coefficient, while the LO coefficient is always taken at full colour. We observe that the LC approximation 
underestimates FC by typically less than one per cent at low $p_T$ and overtakes it by two per cent at high $p_T$. The difference between LC and FC slightly increases towards large $p_T$, most notably in the very forward rapidity bins, but still within the scale uncertainty of the NLO FC.
 \begin{figure}[t]
	\centering
	\includegraphics[width=0.49\textwidth]{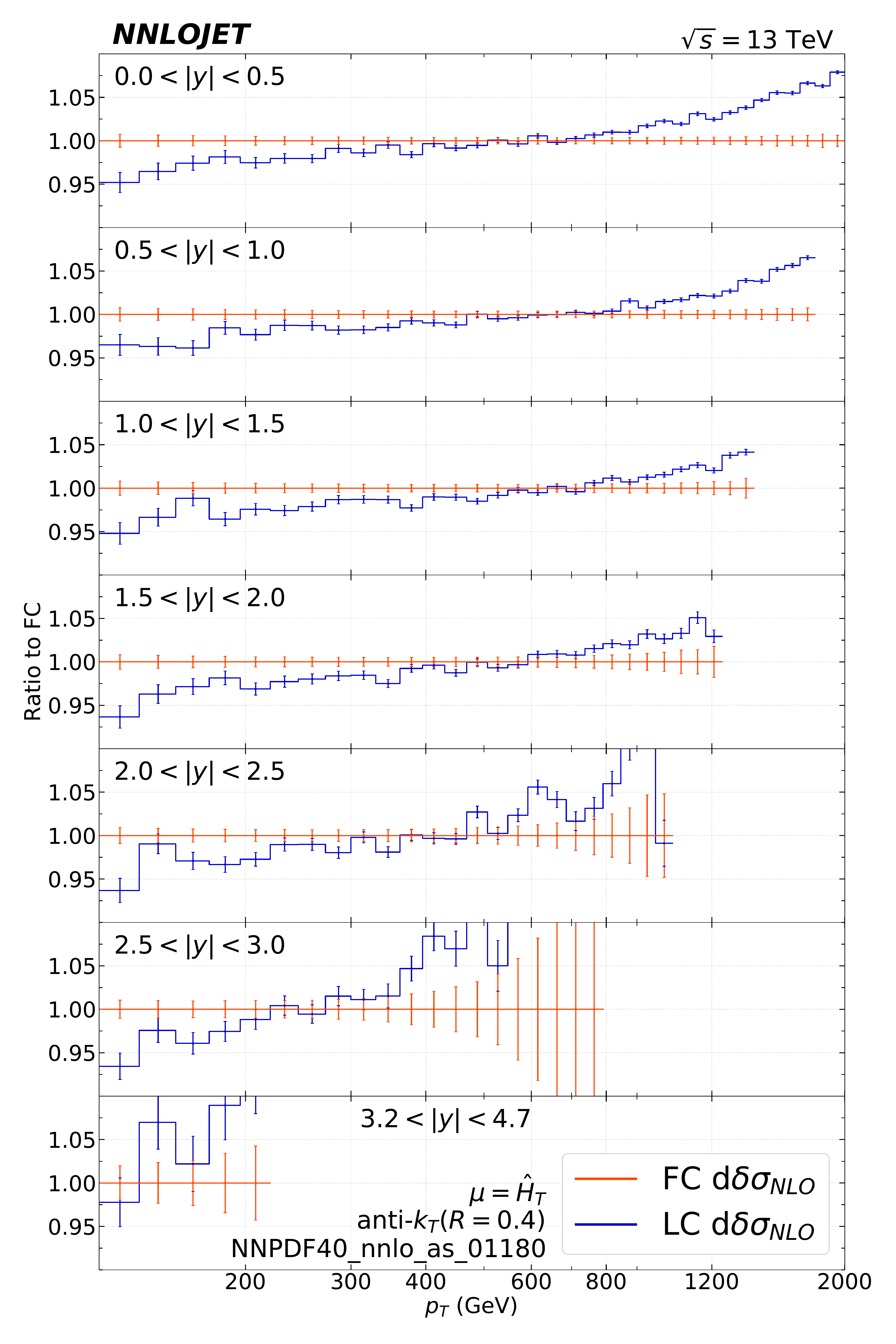}
	\includegraphics[width=0.49\textwidth]{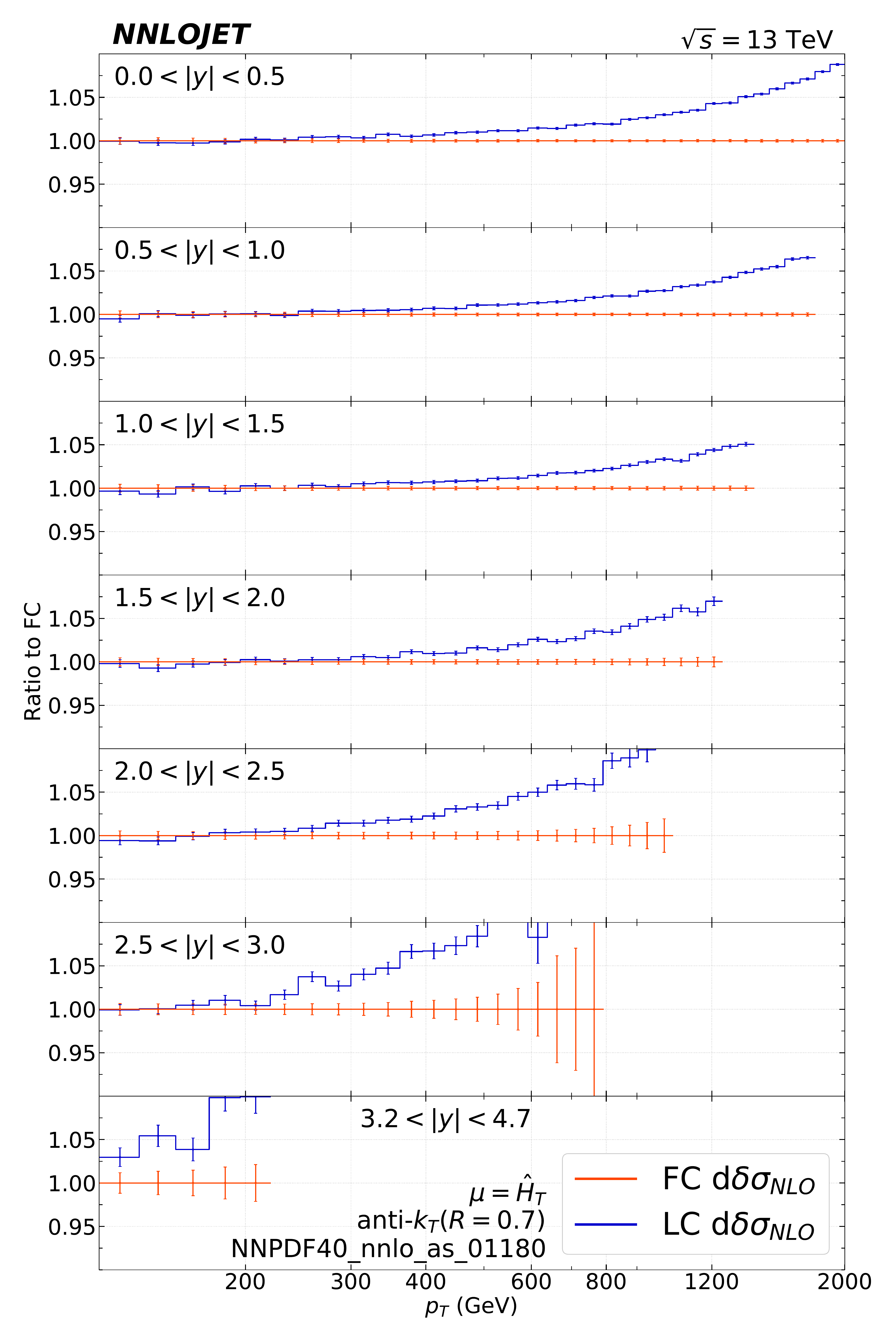}
	\caption{NLO coefficient $\mathrm{d}\delta\sigma_{\mathrm{NLO}}$ at LC (blue) and FC (red) predictions of the $R=0.4$ (left) and $R=0.7$ (right) single jet inclusive cross section as function of $p_T$ for all the considered rapidity slices, normalized to the FC prediction. Ratios are evaluated at the central scale and error bars
	represent numerical integration errors only.}
	\label{fig:CMS13_1j_LCFC_NLO_only}
\end{figure}

In order to quantify the impact of the SLC contributions at NLO more closely, Figure~\ref{fig:CMS13_1j_LCFC_NLO_only} compares the FC and LC
contributions to the NLO coefficient $\mathrm{d}\delta\sigma_{\mathrm{NLO}}$ 
in the single-jet inclusive cross section. It can be seen that 
 SLC contributions of the NLO coefficient range from $+5\%$ at low $p_T$ to $-7\%$ at high $p_T$ of the full colour coefficient for $R=0.4$. For $R=0.7$ the SLC corrections to the LC in the lower $p_T$ and $|y|$ ranges are of the order $1\%$, thus smaller than for the $R=0.4$ jets. The same pattern across the $p_T$ range is observed for both cone sizes however.  
\begin{figure}[t]
	\centering
	\includegraphics[width=0.49\textwidth]{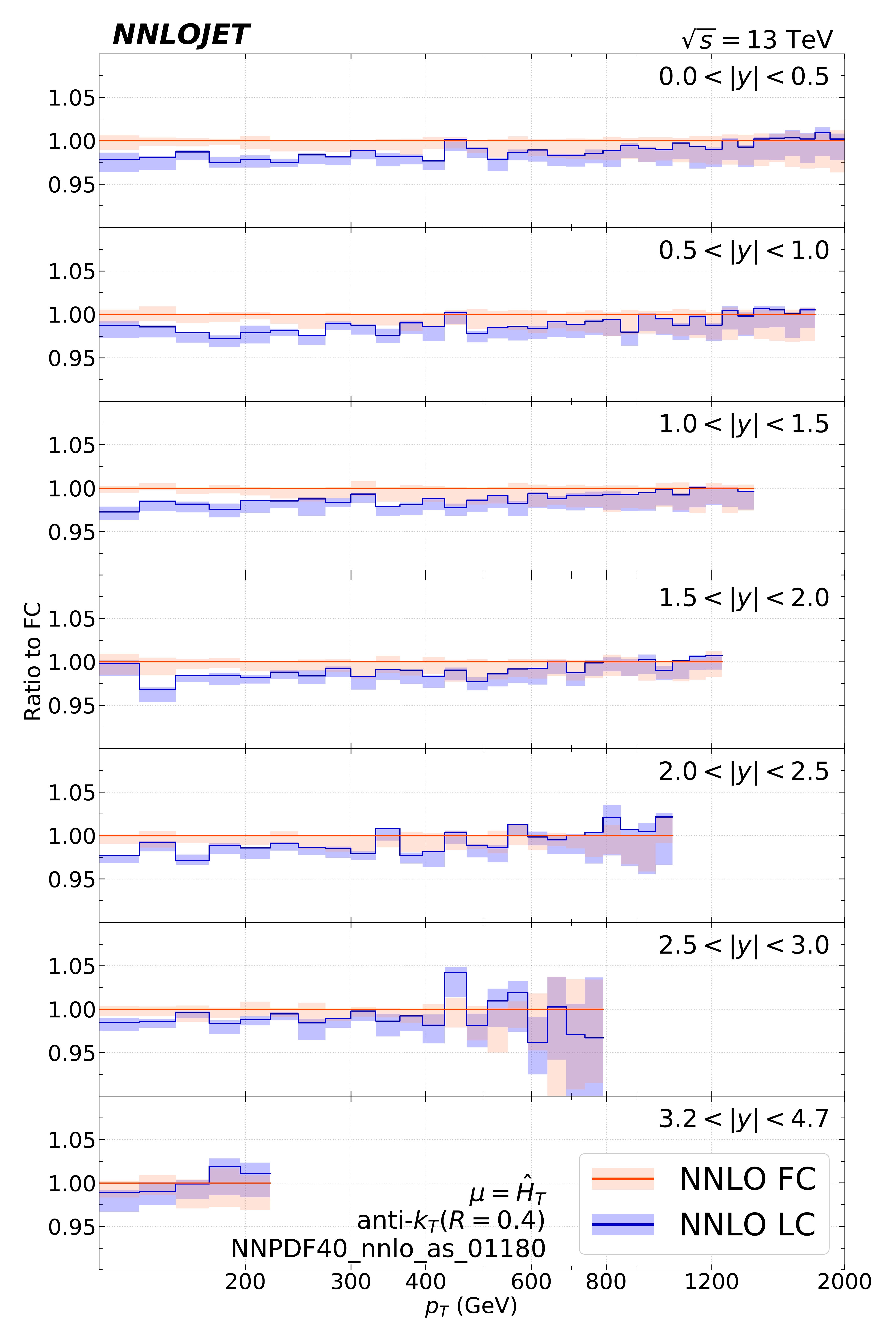}
	\includegraphics[width=0.49\textwidth]{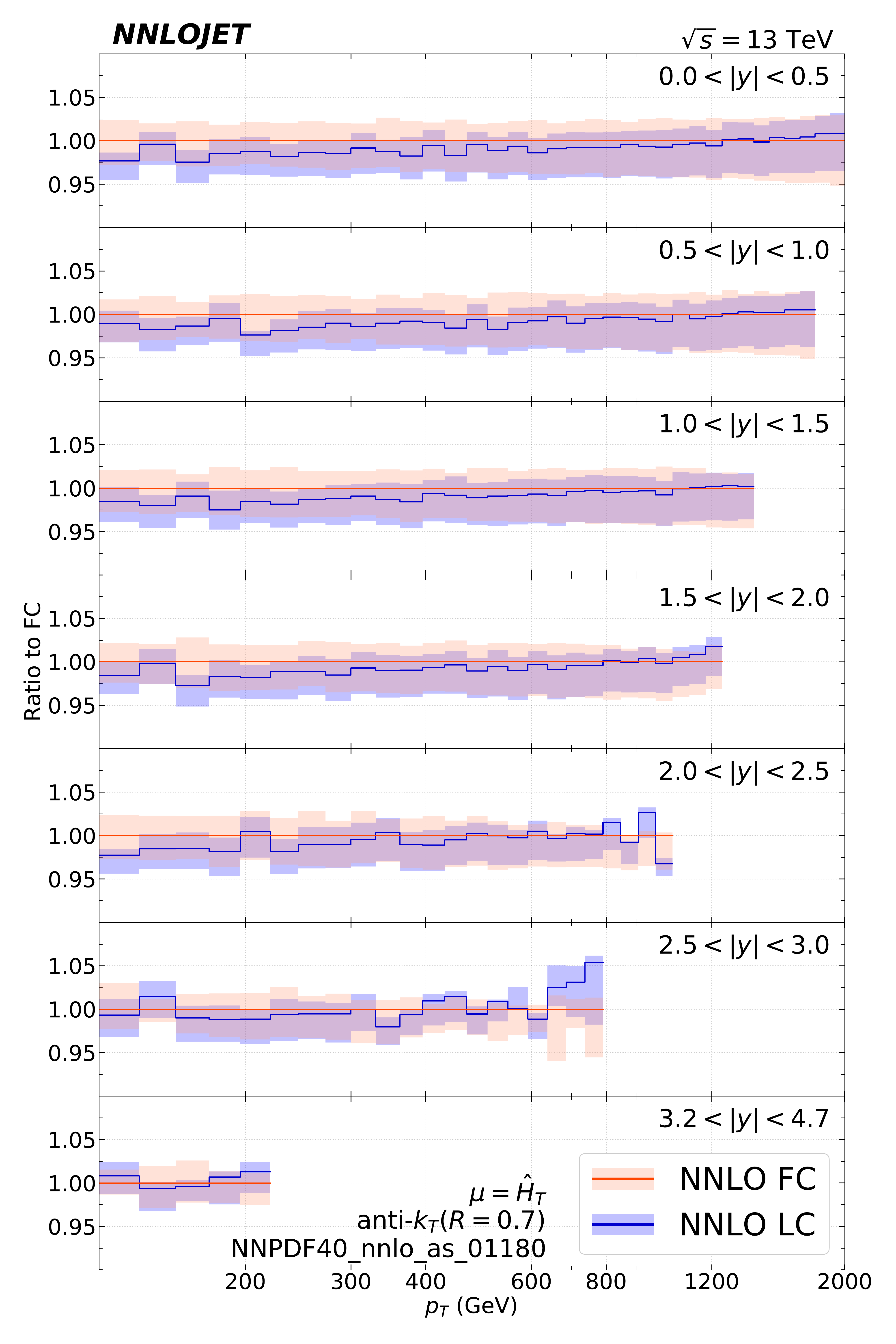}
	\caption{NNLO LC (blue) and FC (red) predictions of the $R=0.4$ (left) and $R=0.7$ (right) single jet inclusive cross section as function of $p_T$ for all the considered rapidity slices, normalized to the FC prediction.}
	\label{fig:CMS13_1j_LCFC_NNLO}
\end{figure}

Given the numerical smallness of the SLC contributions at NLO, it can be expected that their impact at NNLO is equally limited. The good agreement 
of previous results for the $R=0.7$ single-jet inclusive cross section at LC~\cite{Currie:2018xkj} and FC~\cite{Czakon:2019tmo}, which were found 
to be mutually consistent with numerical integration uncertainties (at few per-cent level) lends further support to this expectation. A detailed 
comparison of LC and FC predictions at NNLO is shown in Figure~\ref{fig:CMS13_1j_LCFC_NNLO} for both $R=0.4$ and $R=0.7$ with  
LO and NLO coefficients included at full colour. It can be seen that 
the effect of the SLC contributions is most pronounced at low $p_T$. For $R=0.4$, the SLC corrections enhance the LC predictions by $2$--$3\%$
at low $p_T$ in all rapidity bins. Owing to the very small scale uncertainty (only about 1\% at NNLO) 
at $R=0.4$, this enhancement is significant, and outside 
the previously quoted LC scale uncertainty band. The numerical impact of the SLC corrections is smaller for $R=0.7$, amounting to $1$--$2\%$ at low
$p_T$, and well within the LC scale uncertainty. 
\begin{figure}[t]
	\centering
	\includegraphics[width=0.49\textwidth]{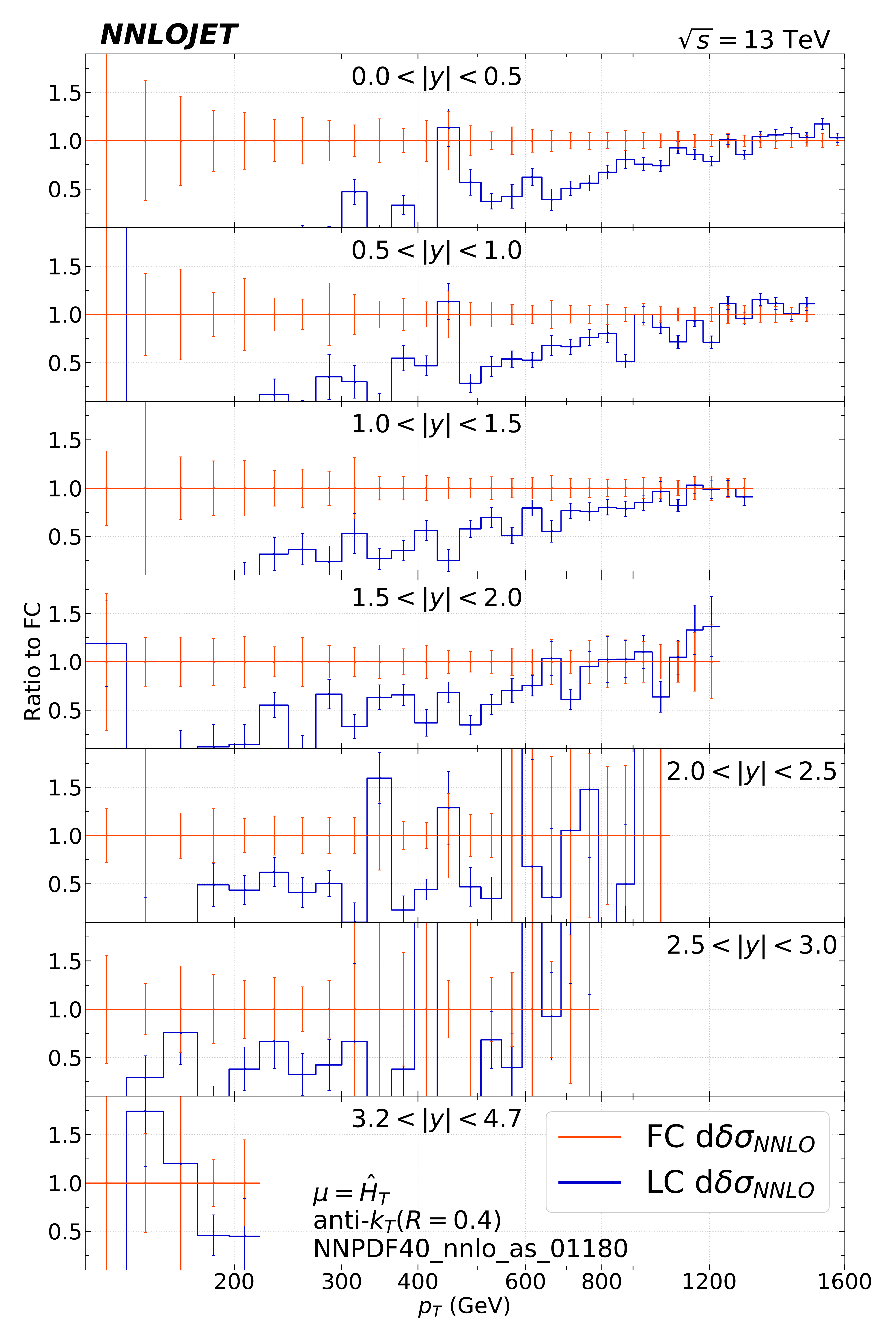}
         \includegraphics[width=0.49\textwidth]{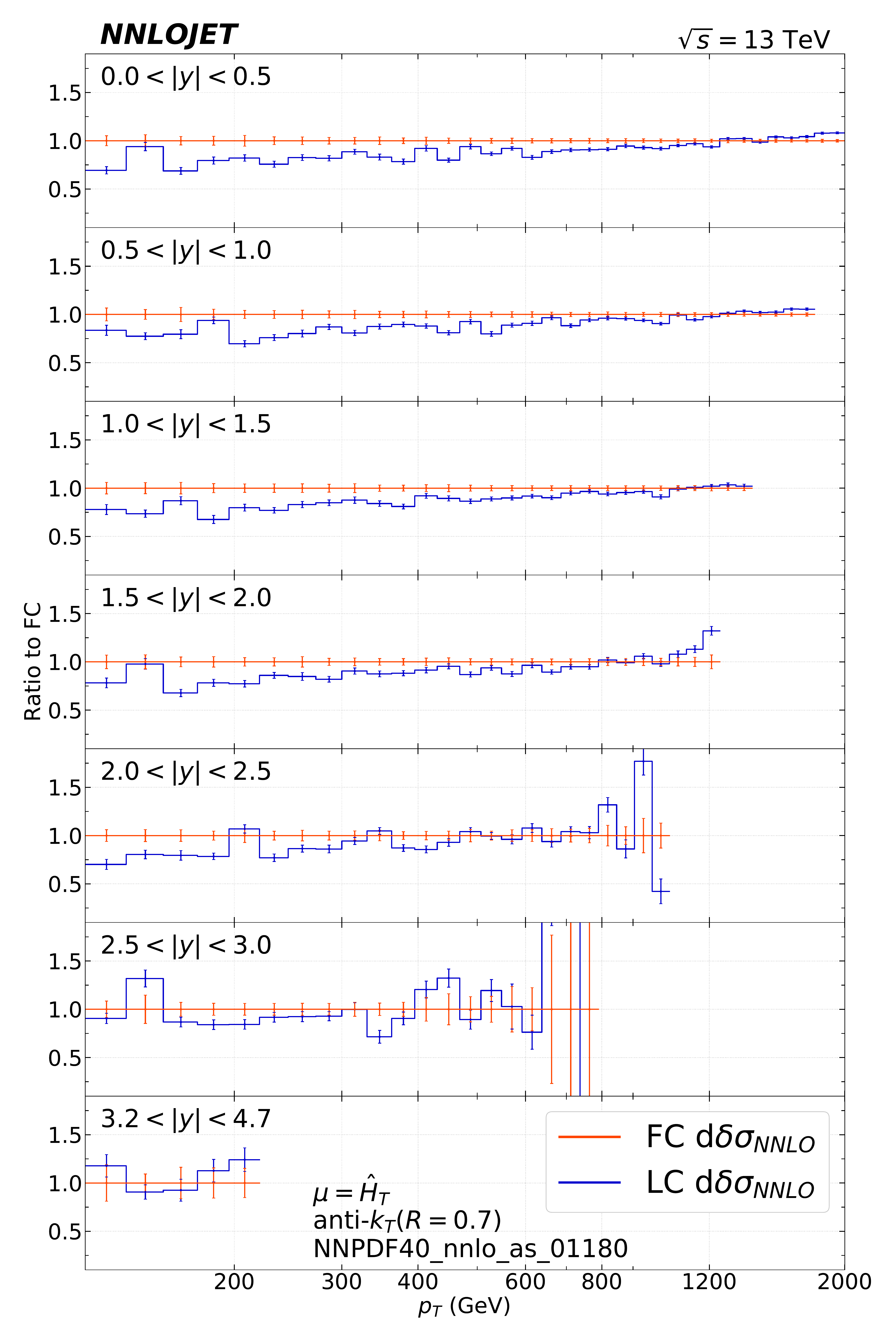}
	\caption{NNLO coefficient $\mathrm{d}\delta\sigma_{\mathrm{NNLO}}$ at LC (blue) and FC (red) predictions of the $R=0.4$ (left) and $R=0.7$ (right) single jet inclusive cross section as function of $p_T$ for all the considered rapidity slices, normalized to the FC prediction. Ratios are evaluated at the central scale and error bars
	represent numerical integration errors only.}
	\label{fig:CMS13_1j_LCFC_NNLO_only}
\end{figure}

The larger relative impact of the SLC effects at $R=0.4$ compared to $R=0.7$ can be understood from the overall
smaller absolute size of the NNLO contribution
for the smaller jet cone size, see Figure~\ref{fig:CMS13_1j_R04} above, which can be related to a decrease of the NNLO LC coefficient
with decreasing $R$. The coefficient is positive for $R=0.7$, but 
crosses zero in the region $R=0.3$--$0.5$~\cite{Currie:2018xkj,Bellm:2019yyh}, depending on the values of $p_T$ and $|y|$.

The resulting smallness of the LC coefficient at $R=0.4$ has to be kept in mind when assessing the relative importance of LC and SLC contributions 
to the NNLO coefficient $\mathrm{d}\delta\sigma_{\mathrm{NNLO}}$ shown in Figure~\ref{fig:CMS13_1j_LCFC_NNLO_only}. For $R=0.7$ (right frame),
 we observe that the LC contribution typically amounts to 80\% or more 
of the FC NNLO coefficient, and that the SLC contribution further diminishes towards large values of $p_T$. In contrast, the relative importance of the SLC 
contributions in the NNLO coefficient is much larger for $R=0.4$, where it clearly dominates over the LC contribution at low $p_T$, becoming of more
moderate impact only towards larger $p_T$. This apparent dominance of SLC contributions 
is however not putting the validity of the LC approximation for this observable into question, since it occurs only in a region where both LC and FC are very small in 
absolute terms, i.e.\ compared to the lower order contributions, as also indicated by the large size of the numerical integration errors
for $R=0.4$ in Figure~\ref{fig:CMS13_1j_LCFC_NNLO_only}. Consequently, the LC/FC ratio is no longer a meaningful quantifier for the validity of the 
LC truncation in this case. 
\begin{figure}[t]
	\centering
			\includegraphics[width=0.49\textwidth]{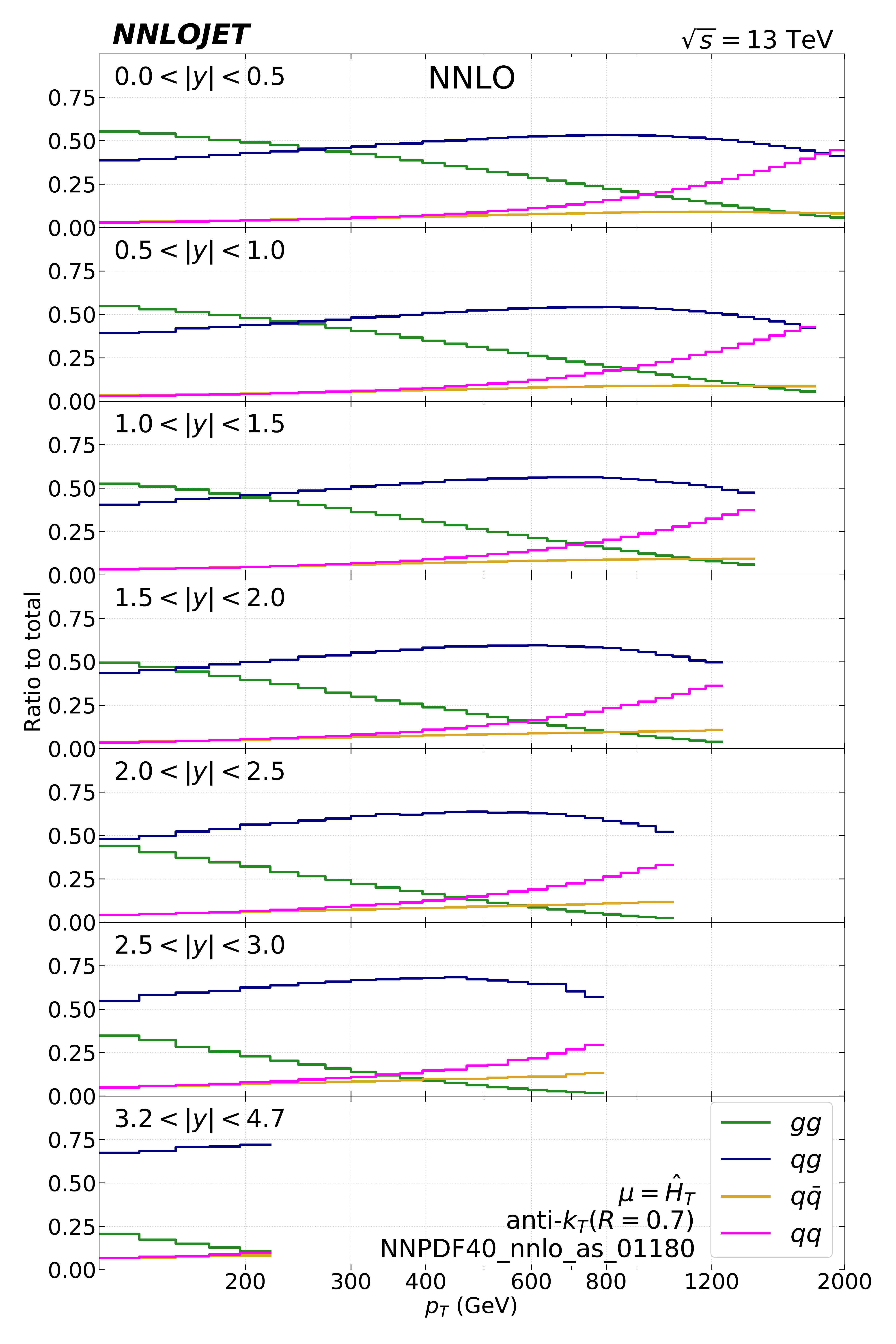}
			\includegraphics[width=0.49\textwidth]{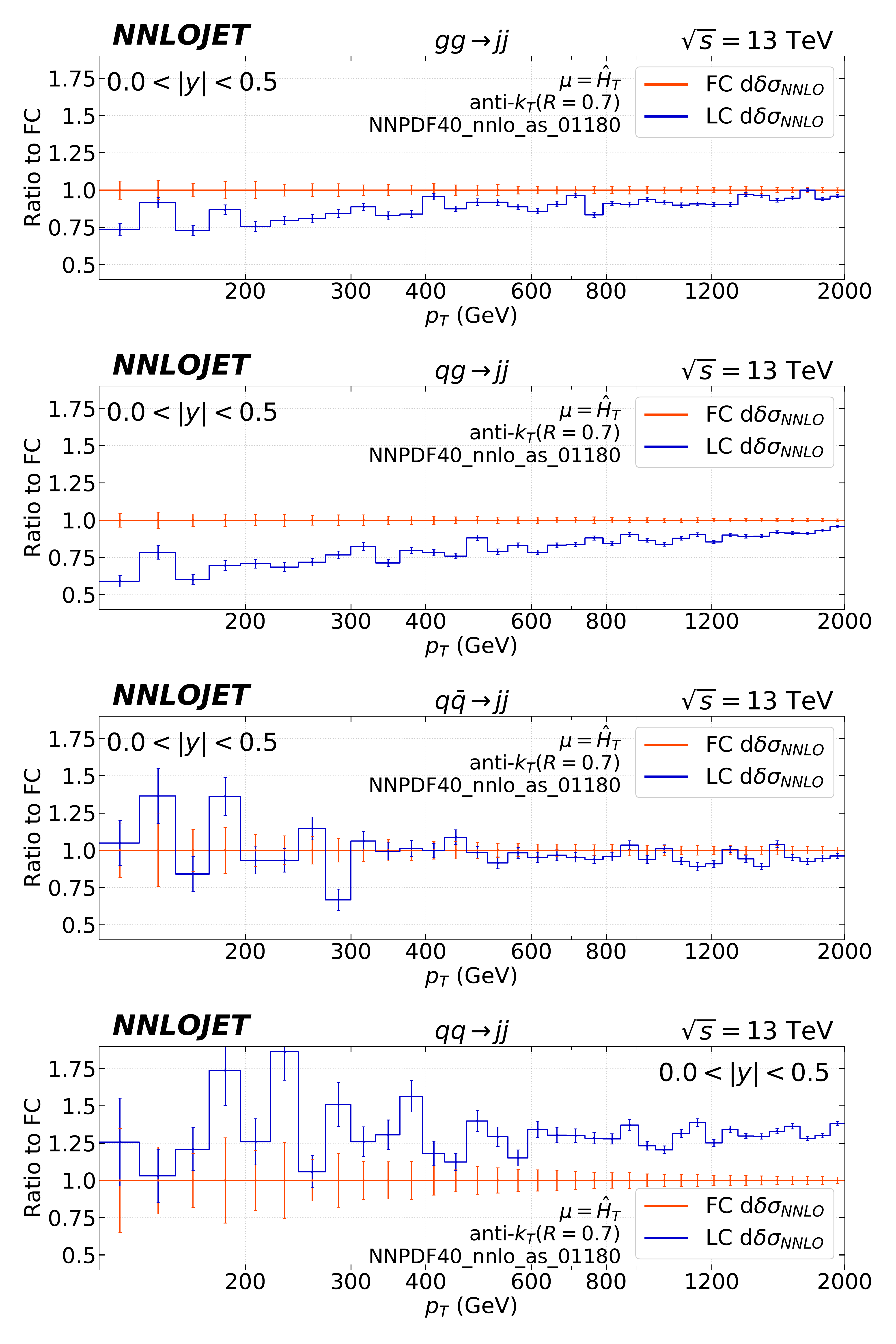} 
	\caption{Relative contribution of each parton-parton initial state to the total single jet inclusive cross section distributions at NNLO (left) and 
	NNLO coefficient $\mathrm{d}\delta\sigma_{\mathrm{NNLO}}$ at LC (blue) and FC (red) predictions for (from top to bottom) the gluon-gluon, quark-gluon, quark-antiquark and quark-quark initiated $R=0.7$ single jet inclusive cross section as function of $p_T$ for the central rapidity slice $0<|y|<0.5$, normalized to the FC prediction (right).}
	\label{fig:CMS13_1j_ChanBreak}
\end{figure}

\subsection{Decomposition into different partonic initial states}  
To further investigate the impact of the SLC contributions at NNLO, we decompose the corrections according to different partonic initial states:
 gluon-gluon ($gg$), quark-gluon ($qg$), quark-antiquark ($q\bar{q}$), quark-quark ($qq$). The different quark flavours are 
 summed over, and the $qg$ and $qq$ initial states are also understood to contain the antiquark-gluon and antiquark-antiquark initiated processes respectively. Given the anomalously small NNLO coefficient 
 at $R=0.4$, this decomposition is performed only for $R=0.7$ where the NNLO coefficient
 is sizeable and displays a sensible hierarchy between LC 
 and SLC contributions. 
 
 The left panel of Figure~\ref{fig:CMS13_1j_ChanBreak} displays the relative importance of the different initial states 
to the single jet inclusive cross section in the different kinematical regions, obtained at NNLO for $R=0.7$. It can be seen that 
$gg$ and $qg$ initial states account for most of the cross section in the central rapidity bins, with $gg$ dominating at low $p_T$ and $qg$ 
taking over at larger $p_T$, where also a sizeable contribution from the $qq$ process is observed. The two foremost bins in rapidity are largely 
dominated by the $qg$ process. Finally, the $q\bar{q}$ process yields only a small contribution throughout the entire kinematical range. 

The LC and FC calculations of the NNLO coefficient  $\mathrm{d}\delta\sigma_{\mathrm{NNLO}}$   are also 
broken down into these four partonic channels. The right panel of 
Figure~\ref{fig:CMS13_1j_ChanBreak} shows the LC/FC ratio for the NNLO contribution in each channel for the $p_T$ distribution in the 
central rapidity bin $0<|y|<0.5$ for $R=0.7$.
The results are in line with the full NNLO coefficient in the central rapidity bin, Figure~\ref{fig:CMS13_1j_LCFC_NNLO_only} (top right). 
The magnitude of SLC contributions in the FC predictions amounts at low $p_T$ 
to  25\% in the $gg$ channel and to up to 40\% in the $qg$ channel. In both channels, the magnitude of the SLC contributions 
decreases with $p_T$, becoming negligible for the highest $p_T$ values.  In the $qq$ channel, SLC contributions are negative, and almost constant 
at $-20$\% of the FC for all $p_T$, thereby leading to partial cancellations 
of the SLC contributions between $qg$ and $qq$ channels at medium $p_T$. 
The $q\bar q$ channel yields the numerically least important contribution to the cross section. It displays 
only minor SLC corrections of less than 10\%, which typically are of the same size or smaller than the numerical integration errors in this channel.

\section{Dijet production}
\label{sec:dijet}

In the following,  we investigate the role of SLC contributions at NNLO in inclusive dijet production at the LHC.
Compared to single-jet inclusive measurements,  dijet final states allow a more differentiated study of the underlying parton-level dynamics since
the underlying Born-level kinematics can at least in principle be fully reconstructed. For each event, only the two jets with the largest transverse momenta 
are used in the kinematical reconstruction of the dijet kinematics, resulting in each event only contributing a single entry into any kinematical dijet distribution (which is in contrast to the single jet inclusive cross section, where each reconstructed jet in an event contributes its own entry into 
the kinematical distributions).  

Experimental measurements of dijet production are commonly performed differential in the dijet invariant mass and
in the rapidities of the jets, often combined into rapidity sums and differences. They  
are then expressed as doubly~\cite{CDF:1999hjm,D0:1998byg,ATLAS:2013jmu,ATLAS:2017ble,CMS:2011nlq}
or triply~\cite{CMS:2017jfq} differential measurements of dijet 
production cross sections. 
The theoretical description of dijet production is identical to that of single-jet inclusive processes, which were outlined 
in detail in Section~\ref{sec:single} above. Specific studies of the NNLO corrections to dijet final states were performed 
in~\cite{Currie:2017eqf} for double differential and in \cite{Gehrmann-DeRidder:2019ibf} for triply differential observables, each time retaining the LC 
contributions only. FC results for dijet production observables have not been computed up to now. 
\begin{figure}[t]
	\centering
	\includegraphics[width=0.49\textwidth]{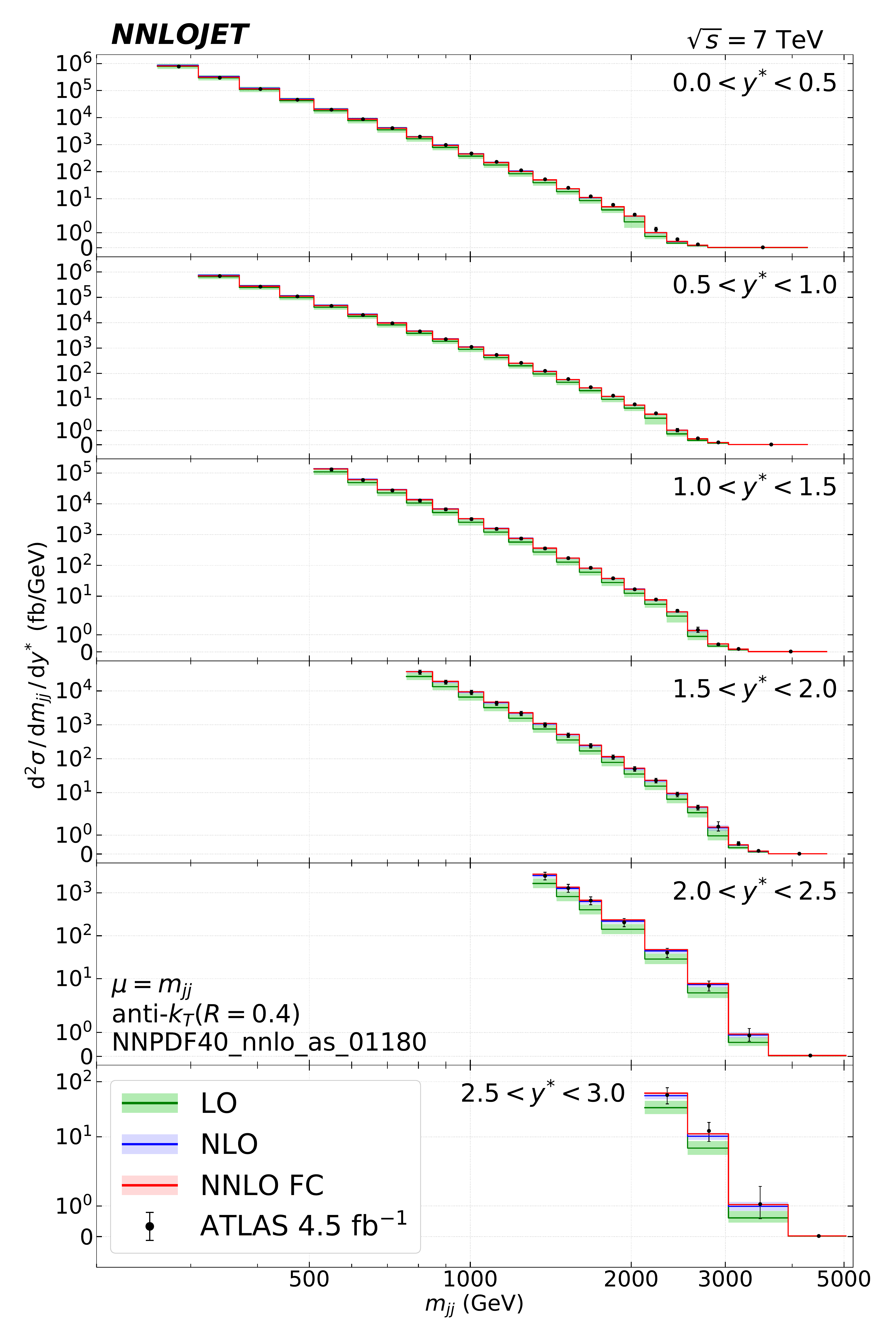} 
	\includegraphics[width=0.49\textwidth]{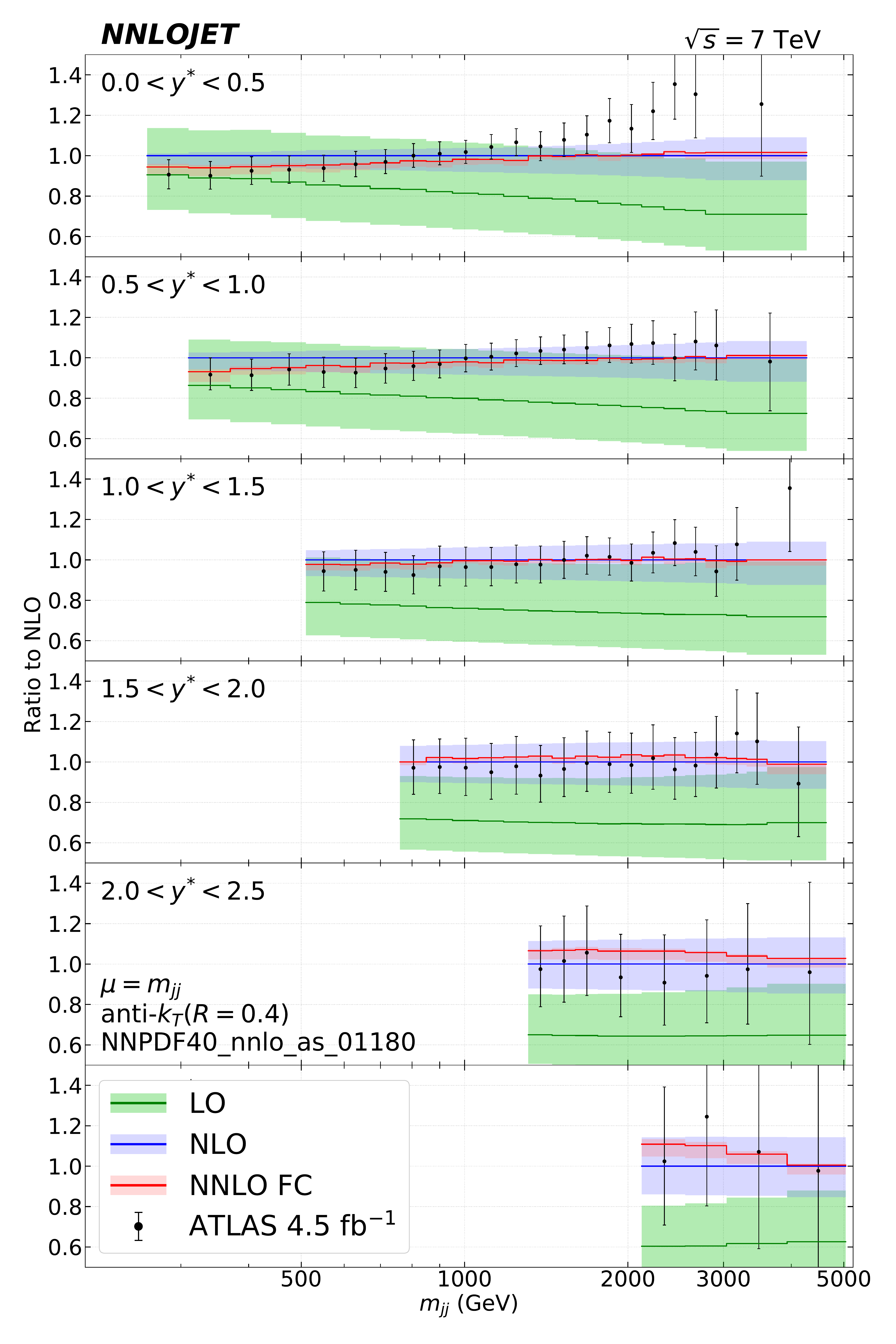}
	\caption{The doubly differential dijet cross section as function of the invariant jet mass $m_{jj}$ for all the considered absolute rapidity difference $y^*$ slices with $R=0.4$ anti-$k_T$ jets compared to the ATLAS 7\,TeV data~\protect\cite{ATLAS:2013jmu}
	in absolute terms (left) and normalized to the NLO prediction (right). }
	\label{fig:ATLAS7_2j}
\end{figure}

\subsection{Calculational setup}
A calculation of dijet production doubly-differential in the dijet mass $m_{jj}$ and the rapidity difference $y^*$ (relating to the Born-level parton-parton scattering angle)
 at NNLO using the LC approximation was performed in~\cite{Currie:2017eqf} and compared with the ATLAS 7\,TeV $4.5\,\mathrm{fb}^{-1}$ 2011 data~\cite{ATLAS:2013jmu}. We redo these predictions at LC and add the newly computed SLC contributions to obtain the FC predictions. The set-up is the same, except that we use NNPDF40\_nnlo\_as\_01180~\cite{Ball:2021leu}
 with $\alpha_s(M_Z) = 0.118$ as PDF instead of the MMHT2014 NNLO PDF sets~\cite{Harland-Lang:2014zoa} for all calculations. LHAPDF~\cite{Buckley:2014ana} is used to evaluate the PDF sets and the strong couping constant $\alpha_s$. Jets are reconstructed using the anti-$k_T$ algorithm~\cite{Cacciari:2008gp} with cone size $R=0.4$ in the rapidity range $|y| < 3$, requiring that the leading jet has a minimum $p_T$ of 100\,GeV and the subleading jet a minimum $p_T$ of 50\,GeV. The double-differential dijet cross sections are presented as distributions in the dijet invariant mass for several rapidity difference slices. The dijet invariant mass and rapidity difference are given by:
\be
m^2_{jj} = (p_{j_1} + p_{j_2})^2, \quad \quad y^* = \frac{1}{2} |y_{j_1} - y_{j_2}|,
\ee
where $p_{j_{1,2}}$ and $y_{j_{1,2}}$ denote respectively the four-momenta and rapidities of the two leading jets.
The rapidity slices have a width of 0.5, ranging from 0.0 to 3.0. The distribution in dijet invariant mass $m_{jj}$ is measured in 21 bins
in the range $260\,\mathrm{GeV}<m_{jj}<4270\,\mathrm{GeV}$.
The theoretical uncertainty band is again obtained from a seven-point scale variation of $\mu_R$ and $\mu_F$ around the central scale $\mu = m_{jj}$.
\begin{figure}[t]
	\centering
	\includegraphics[width=0.49\textwidth]{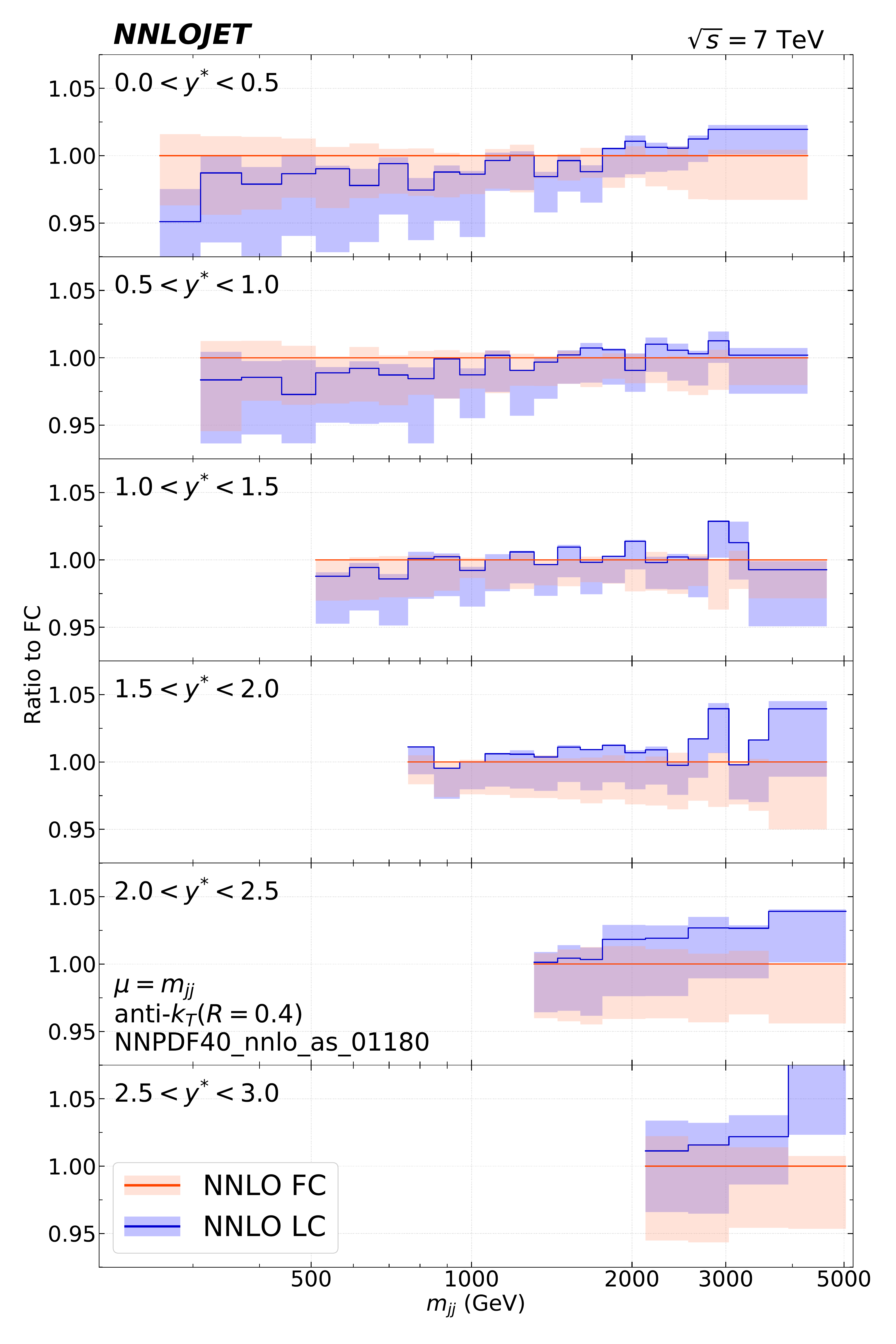}
	\includegraphics[width=0.49\textwidth]{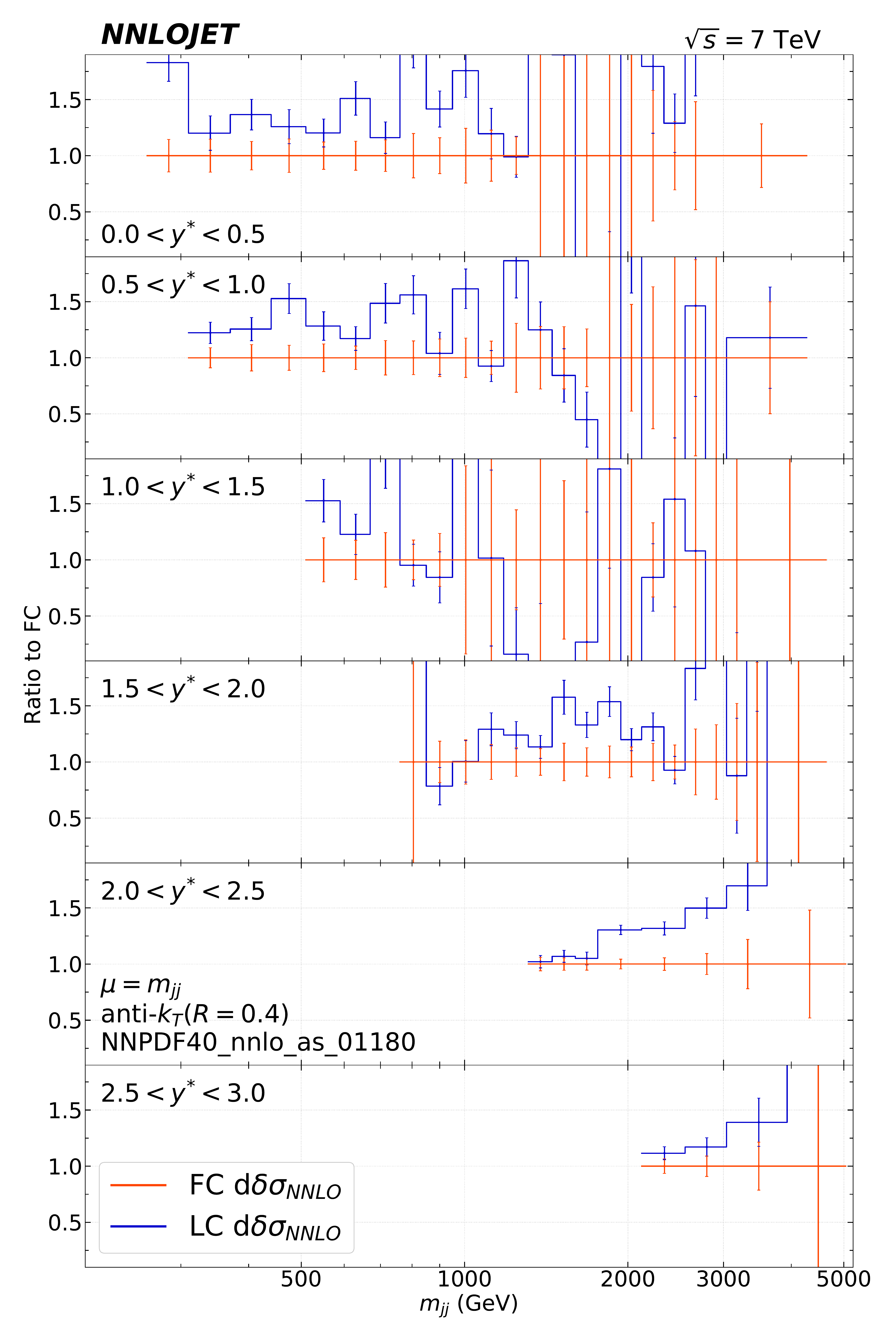}
	\caption{Comparison of LC (blue) and FC (red) predictions of the dijet cross section as function of $m_{jj}$ for all the considered $y^*$ slices, normalized to the FC prediction. Left: NNLO cross section, right: NNLO coefficient. }
	\label{fig:ATLAS7_2j_LCFC_NNLO}
\end{figure}

\subsection{Results and quantification of subleading colour contributions}
Figure~\ref{fig:ATLAS7_2j} displays the LO, NLO and NNLO FC predictions for the dijet doubly differential cross section at 7\,TeV
and compares them
with the ATLAS data~\cite{ATLAS:2013jmu}. For low values of $y^*$, we observe NLO and NNLO corrections to  be only moderate in size, each time 
within the scale uncertainty estimated from the previous perturbative order. In this $y^*$ region, NNLO corrections are negative at low  $m_{jj}$ and 
become nearly vanishing at larger values of  $m_{jj}$. For larger $y^*$, the magnitude of the NLO corrections increases, and NNLO corrections 
become positive throughout the full range of $m_{jj}$. Throughout the full kinematical range, inclusion of NNLO corrections results in a substantial 
decrease of the scale uncertainty to a level of 5\% or below. The experimental data are well described in shape and normalization at NNLO QCD.

The numerical impact of the SLC contributions is illustrated in Figure~\ref{fig:ATLAS7_2j_LCFC_NNLO}. For the NNLO cross section (left frame), 
we observe that FC and LC predictions deviate by no more than 3\% and remain consistent with each other within the NNLO scale uncertainties. 
It is interesting to note that  at low $y^*$ and small $m_{jj}$, the FC predictions are larger than the LC predictions, and that this ordering is reversed at 
larger  $y^*$. Taking account of the overall sign of the NNLO contributions, this implies that inclusion of the SLC contributions leads to a 
decrease of the magnitude of the NNLO corrections throughout the full kinematical range. 

The relative magnitude of the SLC effects in the NNLO coefficient  $\mathrm{d}\delta\sigma_{\mathrm{NNLO}}$   
is shown in the right frame of Figure~\ref{fig:ATLAS7_2j_LCFC_NNLO}. 
The large numerical errors and fluctuations in several of the bins arise from the fact that the FC coefficient (which is used as normalization) 
is close to zero and changing its sign. In the kinematical regions with manifestly non-vanishing FC coefficient, we observe SLC contributions of 
about $-20\%$ throughout, thereby confirming the initial observation of diminishing magnitude of the NNLO corrections from LC to FC.

\section{Triple differential dijet cross section}
\label{sec:3d_2jet}
Lastly, we perform a full colour NNLO calculation of the dijet production cross section triply differential in the average transverse momentum of the two leading jets $p_{T,\mathrm{avg}} = (p_{T,j_1} + p_{T,j_2})/2$, rapidity separation $y^* = |y_{j_1} - y_{j_2}|/2$ and dijet system boost $y_b = |y_{j_1} + y_{j_2}|/2$, improving the LC-only prediction which was first calculated in \cite{Gehrmann-DeRidder:2019ibf}. The triple differential cross section 
has been measured~\cite{CMS:2017jfq} by CMS at 8\,TeV, using an integrated luminosity of 19.7 $\mathrm{fb}^{-1}$,
 as a function of $p_{T,\mathrm{avg}}$ for the following six regions in $(y_b,y^*)$ space:
\ba
y_b \times y^*: &&[0,1] \times [0,1];\quad [0,1] \times [1,2];\quad [0,1] \times [2,3]; \nonumber\\
                &&[1,2] \times [0,1];\quad [1,2] \times [1,2];\quad [2,3] \times [0,1],
\ea
These six regions correspond to different types of event topologies and probe different aspects of the partonic structure of the protons, making this observable ideal for PDF studies and constraints, as explained in detail in~\cite{Gehrmann-DeRidder:2019ibf,CMS:2017jfq}.  
We use the kinematical cuts of the CMS study~\cite{CMS:2017jfq}:
 jets are reconstructed by the anti-$k_T$ algorithm with cone size $R=0.7$ and the fiducial cuts accept events with at least two jets with a maximum absolute rapidity of $|y| \le 5.0$, where the two leading jets in $p_T$ must also have $p_T \ge 50\,\mathrm{GeV}$ and $|y| \le 3.0$. NNPDF4.0 is again used throughout the whole calculation and
 uncertainties on the predictions are determined from a seven-point scale variation around a central scale $m_{jj}$. 

The predictions from \NNLOJET at parton level can be supplemented with nonperturbative (NP) contributions from hadronization and the underlying event. We take the NP contributions derived from parton shower predictions at NLO in \cite{CMS:2017jfq} into account as a multiplicative factor in each bin of the parton level prediction. The resulting 
corrected distributions are indicated as  NNLO$\otimes$NP in the following. 
On top of this, electroweak (EWK) corrections from virtual exchanges of massive $W$ and $Z$ bosons calculated in \cite{Dittmaier:2012kx} are also included as a multiplicative factor, with the resulting distributions labelled as NNLO$\otimes$NP$\otimes$EWK. These EWK multiplicative factors depend only weakly on the PDF set and were determined using NNPDF3.1.

\begin{figure}[t]
	\centering
	\includegraphics[width=1\textwidth]{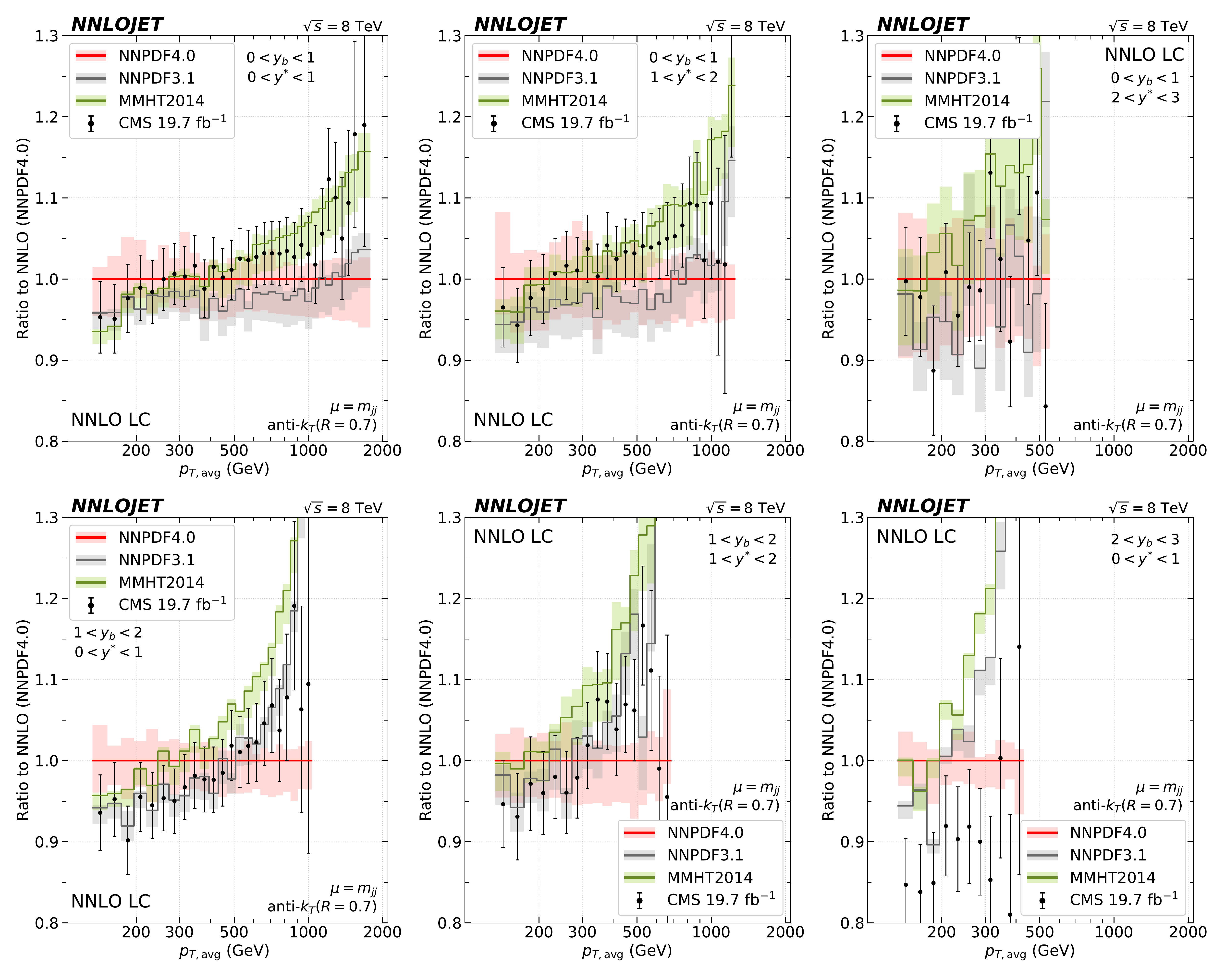} 
	\caption{Comparison of the triple differential dijet distributions at leading colour NNLO$\otimes$NP$\otimes$EW using NNPDF40$\_$nnlo$\_$as$\_$01180 (red), NNPDF31$\_$nnlo$\_$as$\_$0118 (grey) and MMHT2014$\_$nnlo (green) as the PDF. }
	\label{fig:CMS8_3D_PDFcomp_NNLO}
\end{figure}

\subsection{PDF differences}

The CMS 8\,TeV triple differential dijet data~\cite{CMS:2017jfq}  allow to probe the behaviour of the PDFs over a substantial range in parton momentum fraction $x$. Especially the bins with larger
values of $y_b$ correspond to very asymmetric parton-parton collisions, thereby 
directly assessing larger values of proton momentum fraction $x$ than in most single and double differential measurements. 
The impact of various data sets on single inclusive jet production and dijet production at 7\,TeV and 8\,TeV on the determination of the PDFs has been investigated in
detail in~\cite{AbdulKhalek:2020jut} and subsequently in the context of the global NNPDF4.0 PDF fit in~\cite{Ball:2021leu}. In these studies, the CMS 8\,TeV
triple differential jet production data was found to provide the strongest PDF constraints among all jet data set but 
to exercise a substantial pull on the gluon distribution around $x\sim 0.3$. The PDFs 
resulting from a global fit including the CMS 8\,TeV triple-differential measurement as the only jet data set
were consistent with all other jet production 
data but provided a substantially worse description of legacy fixed-target results for deep-inelastic structure functions and Drell-Yan data. 
In view of the resulting deterioration of the overall fit quality, the CMS 8\,TeV triple differential dijet data were discarded (alongside with other, less constraining, dijet data)
in the default NNPDF4.0 PDF fit, which includes only single jet inclusive data. 

Ironically, the newly determined NNPDF4.0 default PDF set~\cite{Ball:2021leu} provides a worse description of the CMS 8\,TeV triple differential dijet data~\cite{CMS:2017jfq}
than the NNPDF3.1~\cite{NNPDF:2017mvq} 
and MMHT2014~\cite{Harland-Lang:2014zoa} PDF sets, which were both determined prior to the CMS measurement. Leading-colour 
NNLO$\otimes$NP$\otimes$EWK predictions obtained with these three PDFs are compared to the CMS data in Figure~\ref{fig:CMS8_3D_PDFcomp_NNLO}, 
illustrating that NNPDF4.0 differs substantially from the other two PDF sets especially in the tail regions. This pattern is already present in
the predictions at  LO and NLO, which are not shown in the figure. 
The best description of the data in the low $y_b$ bins in terms of shape and normalization is provided by the  MMHT2014 PDF set. For the high $y_b$ bins MMHT2014 overshoots the data, while the NNPDF sets describe the data better here, with NNPDF3.1 describing the data especially well in the low $y^*$ bin. 
To maintain consistency with the previous sections, and to adhere to the most up-to-date PDF set, we will nevertheless use the NNPDF4.0 PDFs in the following.

Figure~\ref{fig:CMS8_3D_PDFcomp_NNLO} does not display uncertainties on the predictions due to errors on PDF and $\alpha_s$. Based on a LO evaluation with the NNPDF40$\_$nnlo$\_$as$\_$01180, we expect the PDF uncertainties on the NNLO predictions in Figure~\ref{fig:CMS8_3D_PDFcomp_NNLO} to range from $\pm 0.3\%$ to $0.7\%$ at low $p_{T,\mathrm{avg}}$ to $\pm 2.4\%$ to $8.9\%$ at the highest $p_{T,\mathrm{avg}}$. For a fixed $y_b (y^*)$, the PDF uncertainties increase gradually as $y^* (y_b)$ increases. The combined PDF+$\alpha_s$ uncertainties were estimated at LO with the NNPDF40$\_$nnlo$\_$pdfas set. We find the $\alpha_s$ uncertainties dominant in the low $p_{T,\mathrm{avg}}$ region. The combined uncertainties range from $\pm 1.3\%$ to $1.7\%$ at low $p_{T,\mathrm{avg}}$ to $\pm 2.7\%$ to $9.3\%$ at the highest $p_{T,\mathrm{avg}}$. Taking into account these uncertainty estimates shows that NNPDF40$\_$nnlo$\_$as$\_$01180 is in principle consistent with the CMS triple differential dijet data set within uncertainties, but also highlights the PDF constraints that could be obtained from this data set. 

\begin{figure}[t]
	\centering
	\includegraphics[width=1\textwidth]{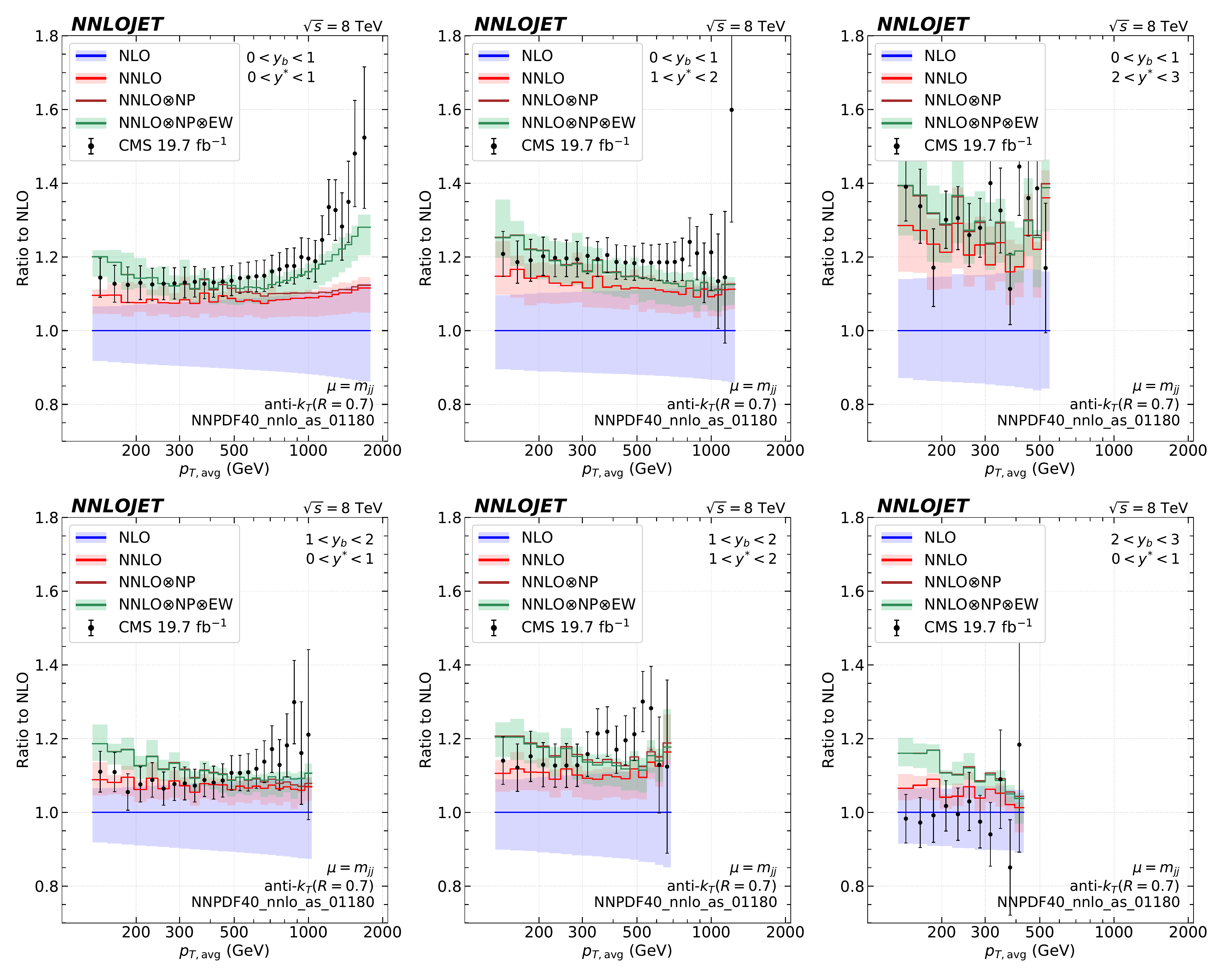} 
	\caption{The triply differential dijet cross section as function of the average transverse momentum $p_{T,\mathrm{avg}}$ for all the considered regions in rapidity separation $y^*$ and dijet system boost $y_b$ space with $R=0.7$ anti-$k_T$ jets compared to the CMS 8\,TeV data~\protect\cite{CMS:2017jfq} normalized to the NLO prediction. }
	\label{fig:CMS8_3D_rat}
\end{figure}

\subsection{Results}

Full-colour NNLO predictions for the triple differential dijet cross section at 8\,TeV are shown in Figure~\ref{fig:CMS8_3D_rat}, where they are compared to
the CMS data~\cite{CMS:2017jfq}. Predictions and data are normalized to the previously available NLO results, and NP and EWK corrections are indicated separately, 
with NNLO$\otimes$NP$\otimes$EWK being the default theory prediction.
Inclusion of the 
NNLO corrections enhances the NLO predictions by about 10\% throughout and  leads to a substantial reduction of theory uncertainties to a residual level of 
$\pm 5\%$ on most distributions. The non-perturbative effects are most pronounced at low $p_{T,\mathrm{avg}}$, while electroweak effects increase towards high 
$p_{T,\mathrm{avg}}$. 
\begin{figure}[t]
	\centering
	\includegraphics[width=1\textwidth]{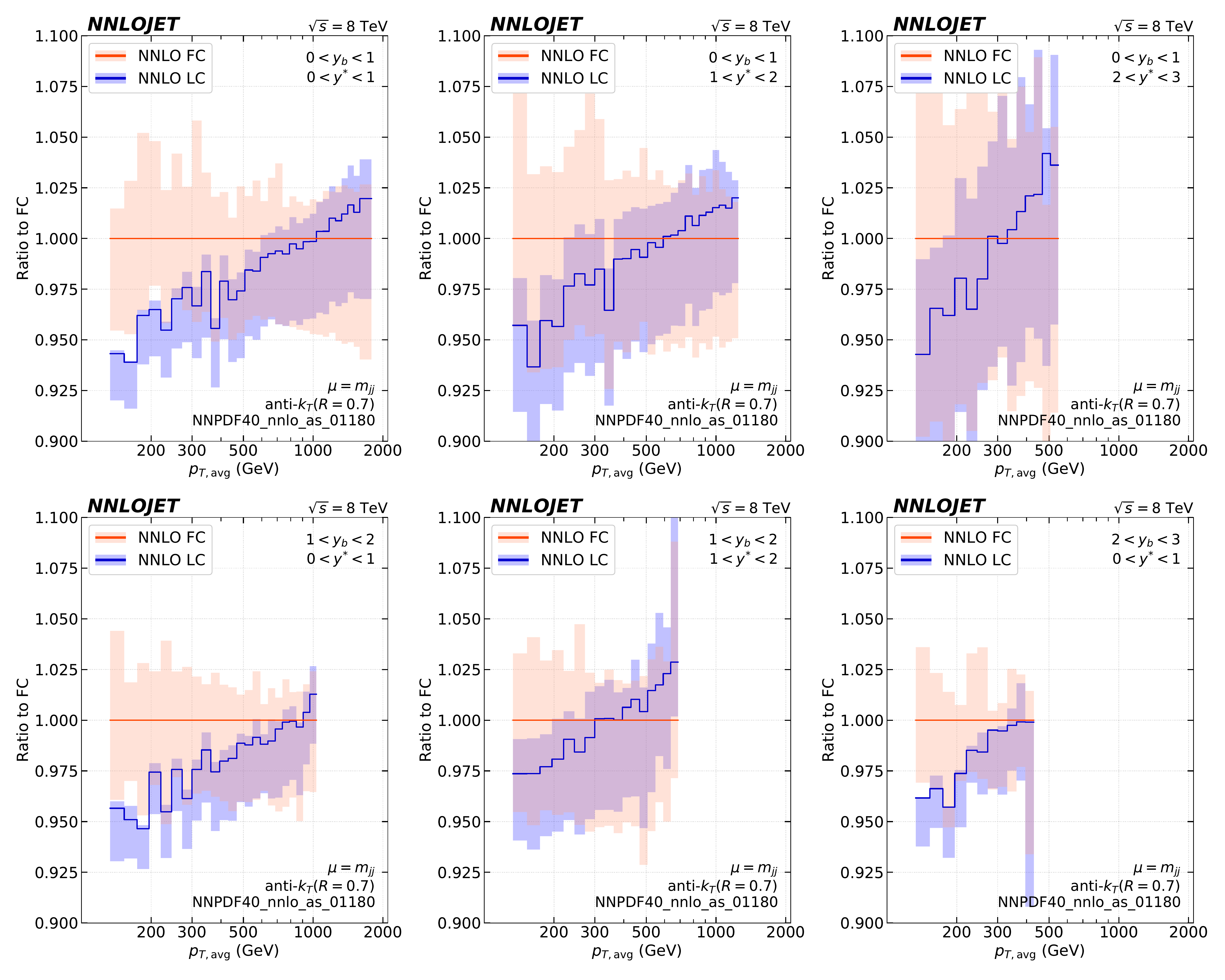} 
	\caption{NNLO LC (blue) and FC (red) predictions of the triple differential dijet distributions normalized to the FC prediction. }
	\label{fig:CMS8_3D_LCFC_NNLO}
\end{figure}

Figure~\ref{fig:CMS8_3D_LCFC_NNLO} compares the FC and LC predictions at NNLO. As before, the LO and NLO coefficients are included in full colour, such that the 
truncation applies only to the NNLO coefficient. In contrast to the single jet inclusive and dijet double differential cross sections, discussed in Sections~\ref{sec:single} and 
  \ref{sec:dijet} above, the SLC contributions are sizable and non-uniform. They typically enhance the LC predictions by about 5\% at low $p_{T,\mathrm{avg}}$, their 
  numerical contribution decreases towards larger values of $p_{T,\mathrm{avg}}$. For central $y_b$ (upper row), the SLC corrections change sign, such that the 
  FC predictions are below the LC predictions for the highest $p_{T,\mathrm{avg}}$ bins. The LC and FC predictions are only marginally consistent with each other within the 
  NNLO scale uncertainty. 
\begin{figure}[t]
	\centering
	\includegraphics[width=1\textwidth]{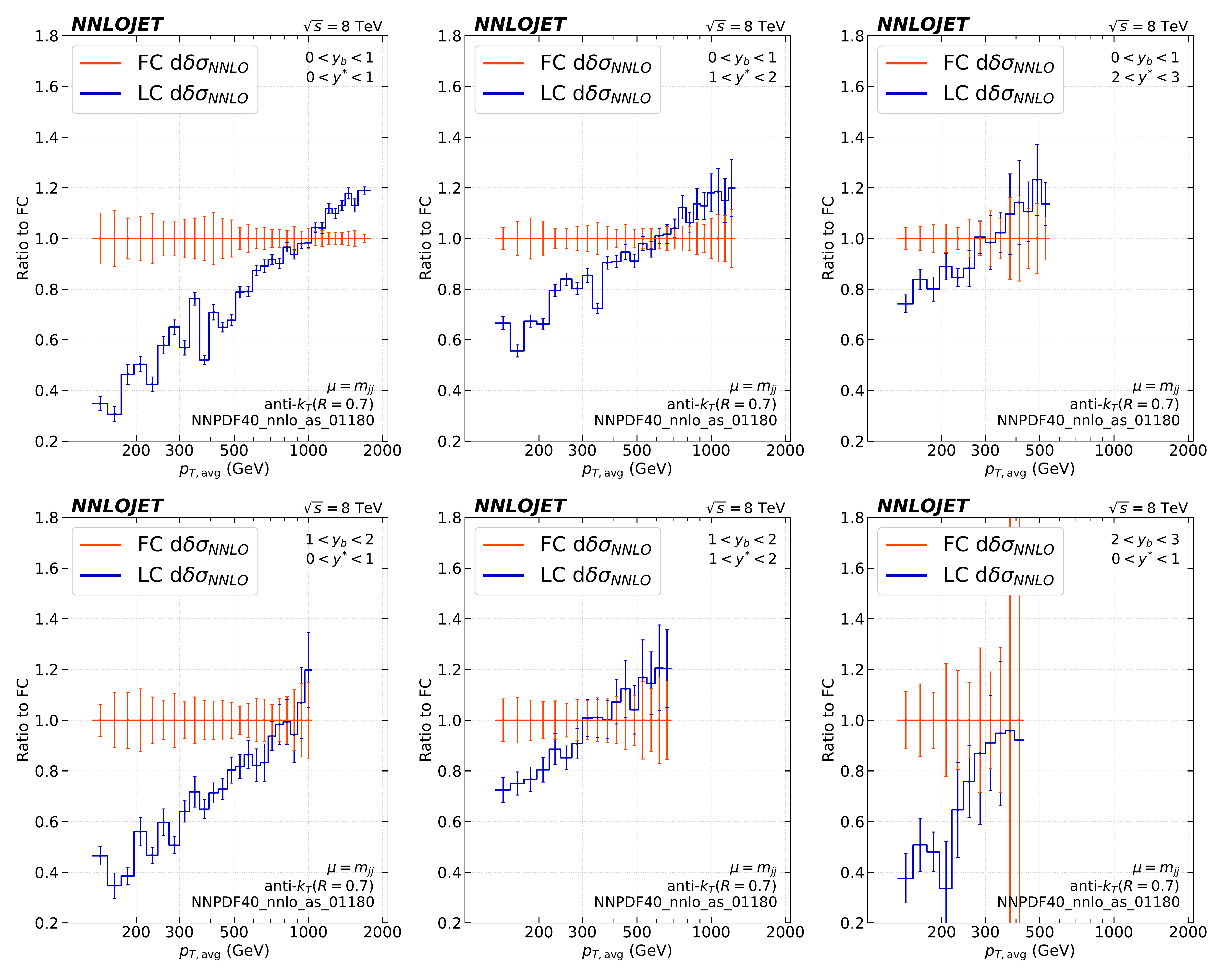} 
	\caption{NNLO coefficient  $\mathrm{d}\delta\sigma_{\mathrm{NNLO}}$  at LC (blue) and FC (red) predictions of the triple differential distributions normalized to the FC prediction. Ratios are evaluated at the central scale and error bars
	represent numerical integration errors only. }
	\label{fig:CMS8_3D_LCFC_NNLO_only}
\end{figure}
  
  The substantial SLC effect on the NNLO coefficient  $\mathrm{d}\delta\sigma_{\mathrm{NNLO}}$  is 
  quantified in Figure~\ref{fig:CMS8_3D_LCFC_NNLO_only}, which compares the LC and FC predictions for this coefficient. The effect is most pronounced at low 
  $p_{T,\mathrm{avg}}$, where LC represents only 40\% of the FC result in in the lowest $y^*$ bins, and typically around 60--70\% in the other bins. With 
  increasing  $p_{T,\mathrm{avg}}$ the LC/FC ratio increases, crossing unity at around $p_{T,\mathrm{avg}}\approx 500\,\mathrm{GeV}$ in most bins and
  reaching 120\% for the largest  $p_{T,\mathrm{avg}}$ values in some bins. 
 
In direct comparison with the results from Section~\ref{sec:dijet} on the double differential dijet 
cross sections at $R=0.4$, we find that numerical impact of the SLC corrections on the NNLO coefficient is considerably larger in the triple differential dijet cross sections 
at $R=0.7$. In particular, large positive SLC corrections at low $p_{T,\mathrm{avg}}$ are observed for all values of $y^*$ in the triple differential distributions, while 
they are concentrated only at low $y^*$ in the double differential distributions. Phenomenologically, this effect is further aggravated by the jet size dependence of the NNLO 
corrections that was already observed for the single jet inclusive distributions in Section~\ref{sec:single}: the NNLO corrections are very close to zero for $R=0.4$, while they are typically positive and of the order of +10\% for $R=0.7$. Consequently, in the triple differential case at $R=0.7$, the SLC corrections have a 
substantial absolute 
impact on the full NNLO prediction, while showing little to no impact for $R=0.4$ in the double differential case, despite yielding a sizeable relative contribution to 
the NNLO coefficient in both cases. 

\begin{figure}[t]
	\centering
	\includegraphics[width=1\textwidth]{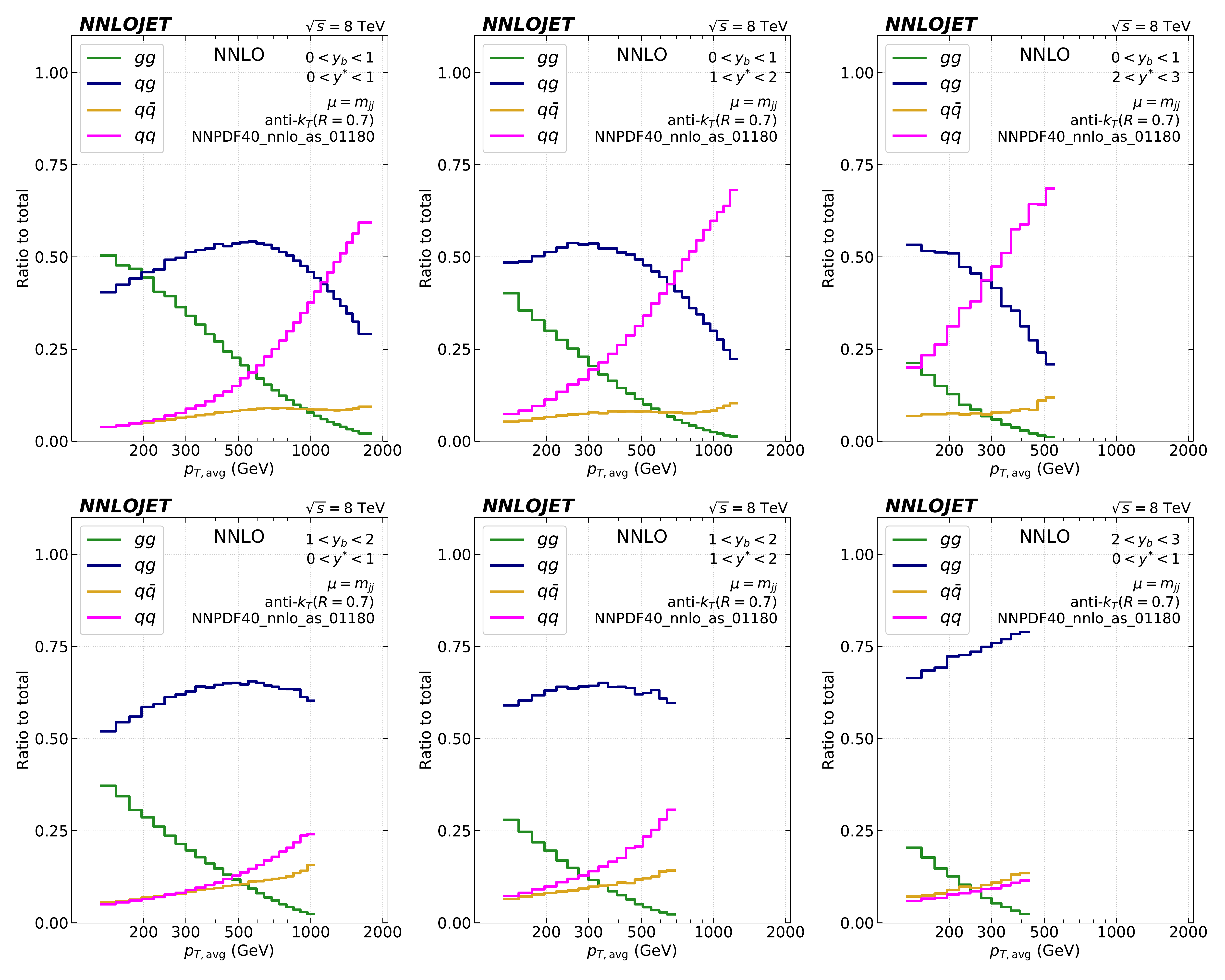} 
	\caption{Relative contribution of each parton-parton initial state to the total dijet cross section distributions in the different bins in $(y_b,y^*)$, determined 
	at  NNLO.}
	\label{fig:CMS8_3D_ChanBreak}
\end{figure}
\subsection{Decomposition into different partonic initial states}
In view of using data on triple differential dijet production to constrain PDFs, it is important to determine the magnitude of subleading colour contributions for 
different partonic initial states, also taking into account the relative importance the different initial states across the kinematical range of the measurement. 
Figure~\ref{fig:CMS8_3D_ChanBreak} displays the breakdown of the triple differential dijet cross section into different partonic channels. 
It can be seen that at central $y_b$ (upper row), gluon-gluon and quark-gluon initial states contribute about equally at low $p_{T,\mathrm{avg}}$. The 
relative importance of the quark-gluon process remains nearly constant for most of the range in $p_{T,\mathrm{avg}}$ (decreasing only at the highest values), 
while 
the contribution from gluon-gluon initial states is rapidly decreasing with $p_{T,\mathrm{avg}}$, with quark-quark processes increasing accordingly to 
become dominant at large $p_{T,\mathrm{avg}}$. The quark-antiquark process remains subdominant throughout the whole distribution. 
For larger values of $y_b$ (lower row), which correspond to asymmetric collisions, quark-gluon initiated processes
dominate throughout the entire  $p_{T,\mathrm{avg}}$ range, typically accounting for more than half of the cross section. The remainder of the cross section is 
given by gluon-gluon initial states at low $p_{T,\mathrm{avg}}$ and quark-quark and quark-antiquark initial states at larger $p_{T,\mathrm{avg}}$. In contrast to 
the central $y_b$ range (upper row), quark-quark and quark-antiquark initial states contribute about equal amounts for the bins with larger $y_b$ (lower row). 
The different qualitative behaviour can be understood as follows: at central $y_b$, both incoming partons carry comparable momentum fractions, such that the 
$p_{T,\mathrm{avg}}$ distribution turns from all-gluon and sea-quark processes at low values to valence-quark scattering at high values. The more 
asymmetric collisions at larger $y_b$ always involve one low-momentum parton (a gluon or a sea quark/antiquark) and one parton at larger momentum (which can be 
a gluon in the low $p_{T,\mathrm{avg}}$ range, but is almost certainly a valence quark for events at larger $p_{T,\mathrm{avg}}$). 
\begin{figure}[t]
	\centering
	\includegraphics[width=1\textwidth]{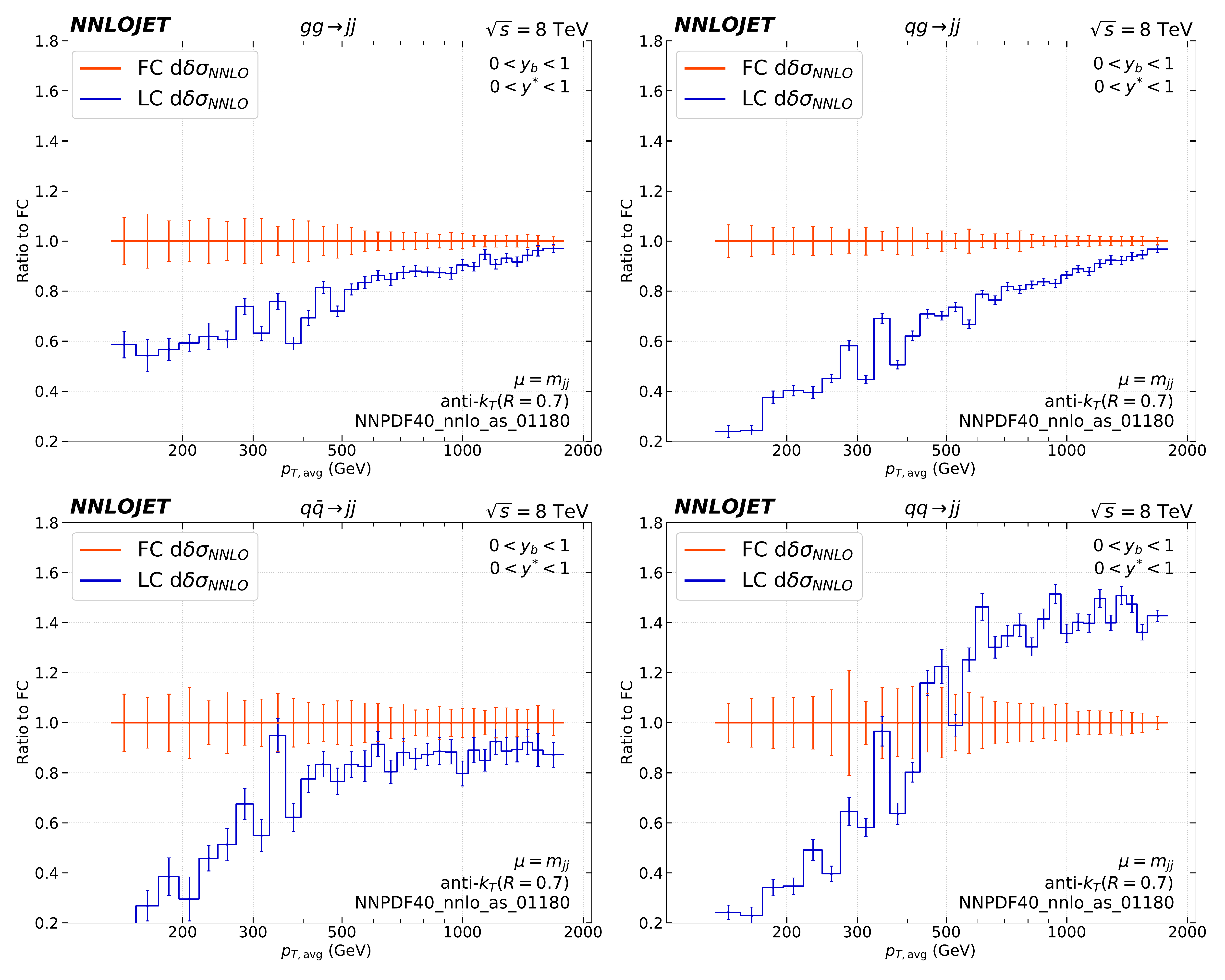} 
	\caption{NNLO coefficient  $\mathrm{d}\delta\sigma_{\mathrm{NNLO}}$  at LC (blue) and FC (red) for the gluon-gluon, quark-gluon, quark-antiquark and quark-quark initiated dijet cross section in the $(y_b,y^*) = ([0;1],[0;1])$ bin. Ratios are evaluated at the central scale and error bars
	represent numerical integration errors only.}
	\label{fig:CMS8_3D_LCFC_ChanBreak_yb0_ys0_NNLO_only}
\end{figure}

The impact of SLC contributions on the 
NNLO coefficient  $\mathrm{d}\delta\sigma_{\mathrm{NNLO}}$  for the different partonic channels is exemplified in Figure~\ref{fig:CMS8_3D_LCFC_ChanBreak_yb0_ys0_NNLO_only} for the first bin 
with $(y_b,\, y^*) = ([0,1],\,[0,1])$. At low $p_{T,\mathrm{avg}}$ it can be seen
for all channels that the SLC contributions are positive and comparable in magnitude to the LC contributions. Towards larger values of $p_{T,\mathrm{avg}}$,
the relative importance of the SLC contributions diminishes, levelling below a level of $+10\%$ of the NNLO coefficient above $p_{T,\mathrm{avg}}\approx 300\,\mathrm{GeV}$ for all
channels except the quark-quark scattering, where the SLC corrections become negative at $-30\%$ for large  $p_{T,\mathrm{avg}}$. It should be noted that this
change occurs precisely in the kinematical range where the quark-quark subprocess starts to yield a dominant contribution to the triple differential cross dijet cross section.  

Given the numerical impact of the 
SLC corrections and their variation between the different partonic channels, it 
appears mandatory to properly account for these corrections in future NNLO PDF fits aiming to include the experimental information from triple differential dijet production. 
In particular, it will be worthwhile to investigate whether the observed tension~\cite{Ball:2021leu} of PDFs obtained including the CMS 8\,TeV triple differential dijet
data set~\cite{CMS:2017jfq} with other data sets persists or is eased when including the SLC corrections at NNLO.

\section{Conclusions}
\label{sec:conc}
In this paper, we presented a new calculation of  the NNLO corrections to jet production observables at hadron colliders, using the antenna subtraction method 
to handle infrared singularities in the real radiation contributions, implemented in the \NNLOJET parton-level event generator code. Previous \NNLOJET results 
for these observables~\cite{Currie:2016bfm,Currie:2017eqf,Gehrmann-DeRidder:2019ibf} and phenomenological studies derived from 
them~\cite{Currie:2018xkj,AbdulKhalek:2020jut} retained only the leading terms in an expansion in $N_c$ and $n_f$ in the NNLO coefficients. Our newly derived 
results improve upon these leading colour approximations by computing the full-colour expressions, thus being exact in the $N_c$ and $n_f$ dependence at NNLO. 

We assessed the numerical impact of the newly derived subleading colour contributions in detail for single jet inclusive and dijet production observables,
taking the  kinematical definitions used in the 
CMS single jet inclusive measurements at $\sqrt{s}=13\,\mathrm{TeV}$~\cite{CMS:2016jip}, the
ATLAS 7\,TeV   doubly differential dijet cross section~\cite{ATLAS:2013jmu} and the
CMS 8\,TeV triple differential dijet cross section~\cite{CMS:2017jfq}
as reference points. These measurements span a diversity of kinematical settings and jet cone sizes. 

For the single jet inclusive cross section at $R=0.4$ and $R=0.7$, we find the subleading colour contributions to be of very small impact on the NNLO predictions, usually 
within the previously quoted theory uncertainties. Our findings agree with the results of a 
full-colour calculation~\cite{Czakon:2019tmo} of this observable at $R=0.7$ that used a different infrared subtraction method. 

For dijet cross sections, we observe a 
more differentiated pattern in the subleading colour contributions. 
While these contributions are observed to have only a small effect on the NNLO predictions 
for double differential dijet distributions at $R=0.4$, their effect is more pronounced in triply differential distributions at $R=0.7$. Given that the latter observable 
provides potentially important constraints on the parton distributions in the proton, thereby displaying a tension with data sets from other processes in a global 
PDF fit, the inclusion of the newly derived full-colour NNLO results could have a potential impact on the precision understanding of PDFs from LHC jet data. 
The magnitude of the subleading colour contributions to the NNLO coefficients is non-uniform in the kinematical variables and in the different parton-level initial states,
with cancellations occurring among different partonic channels. 

By comparing the numerical impact of the subleading colour contributions on the different observables, some pattern appears to emerge. All observables receive 
large NLO corrections of comparable size, and the subleading colour effects in the NLO coefficients are small, typically in the range below ten per cent, as expected from a 
naive $1/N_c^2$ power counting. The magnitude of the NNLO corrections varies substantially between different observables. For those observables where the 
NNLO/NLO $K$-factors are comparable in magnitude to the NLO/LO $K$-factors, we do indeed observe small subleading colour effects that are in line with the naive power counting expectations. In contrast, those observables that display smaller NNLO/NLO $K$-factors typically receive larger subleading colour contributions to their 
NNLO coefficients. We can only speculate about the underlying mechanism. Some further insight may be gained by noting that leading logarithmic effects from 
multiple resolved real radiation only contribute at the leading colour level to each order in the perturbation expansion, thereby populating some parts of the final state phase 
space more strongly than others. Subleading colour contributions do not receive this enhancement, and could thus potentially be more uniform in phase space. 
Observables that are not sufficiently inclusive on this extra radiation (for example due to a small jet cone size or due to multiple kinematical constraints) could thus 
display anomalously small leading colour contributions at NNLO, resulting in an enhanced numerical importance of subleading colour effects.

\section*{Acknowledgements}

The authors thank Juan Cruz-Martinez, James Currie, Rhorry Gauld, Aude Gehrmann-De Ridder,
Marius H\"ofer, Imre Majer, Matteo Marcoli, Thomas Morgan, Jan Niehues, Jo\~ao Pires, Adrian Rodriguez Garcia, 
Robin Sch\"urmann, Giovanni Stagnitto, Duncan Walker, Steven Wells and James Whitehead for useful 
discussions and their many contributions to the \NNLOJET code.
This work has received funding from the Swiss National Science Foundation (SNF) under contract 200020-204200, from the European Research Council (ERC) under the European Union's Horizon 2020 research and innovation programme grant agreement 101019620 (ERC Advanced Grant TOPUP), from the UK Science and Technology Facilities Council (STFC) through grant ST/T001011/1 and from the Deutsche Forschungsgemeinschaft (DFG, German Research Foundation) under grant 396021762-TRR 257.


\bibliographystyle{JHEP}
\bibliography{jetfc_phen}

\end{document}